\documentclass[11pt]{article}
\usepackage{import}

\usepackage[dvipsnames]{xcolor}

\pdfoutput=1
\usepackage{jheppub} 
\usepackage{amsmath,amssymb,epsfig,amsfonts}
\usepackage{epsf}
\usepackage{makeidx}
\usepackage{amsmath}
\usepackage{graphicx}
\usepackage{bbm}
\usepackage{amssymb}
\usepackage{enumitem}
\usepackage{booktabs}
\usepackage{verbatim} 
\usepackage{soul}
\usepackage{mathrsfs}
\usepackage{bm}
\usepackage{subcaption}
 \usepackage{slashed}
    \usepackage{verbatim} %for comments
    \usepackage{float}
    \restylefloat{table}
    \usepackage{pdflscape}
    \usepackage{tikz}
    \usepackage{pifont}
\usepackage{array}
    \usepackage{youngtab}
    \usepackage{multirow}
    \usepackage{ulem}
    \usepackage{cleveref}
    \usepackage{tensor}
    \usepackage{todonotes}
    \usepackage{mathrsfs}
    \usepackage{changepage}
    \usepackage{anyfontsize}
   \usepackage{caption}
\usepackage{orcidlink} 
\usepackage{ytableau}
\usepackage{subcaption}

\makeatletter

\usepackage{tikz-3dplot}

\newcommand{\be}{\begin{equation}}
\newcommand{\ee}{\end{equation}}
\newcommand{\dd}{\mathrm{d}}

\newcommand{\me}{\mathrm{e}}
\newcommand{\ii}{\mathrm{i}}
\newcommand{\vol}{\mathrm{vol}}

\newcommand{\del}{\partial}
\newcommand{\ls}{\ell_s}

\newcommand{\nn}{\nonumber}

\newcommand{\hook}{\mathbin{\rule[.2ex]{.4em}{.03em}\rule[.2ex]{.03em}{.9ex}}}

\newcommand{\Rmnum}[1]{\expandafter\@slowromancap\romannumeral #1@}
\makeatother

\def\bea{\begin{eqnarray}}
\def\eea{\end{eqnarray}}

\newcommand{\diff}{\mathrm{d}}
\newcommand{\R}{\mathbb{R}}
\newcommand{\Z}{\mathbb{Z}}
\newcommand{\C}{\mathbb{C}}

\newcommand{\y}{\Upsilon}

\newcommand{\Fm}{F^\mathrm{RR}}

\newcommand{\ma}{m}
\newcommand{\pt}{\mathrm{p}}

\definecolor{asparagus}{rgb}{0.53, 0.66, 0.42}
\definecolor{bittersweet}{rgb}{1.0, 0.44, 0.37}

\makeatother

\begin{document}

% format
\baselineskip=18pt  % a la harvmac
\numberwithin{equation}{section}  % make eq labels (sec.num)
\allowdisplaybreaks  % allow page breaks in displayed eqs

%%%%%%%%%%%%%%%%%%%%%%%%%%%%%%%%%%%%%%%%%%%
%%%        TITLE BEGINS HERE
%%%%%%%%%%%%%%%%%%%%%%%%%%%%%%%%%%%%%%%%%%%

\thispagestyle{empty}

%title, authors, affiliation
\vspace*{1cm} 
\begin{center}
{\fontsize{18pt}{23pt}\selectfont\textbf{Ten-dimensional localization}\vspace{10mm}}
 
 \renewcommand{\thefootnote}{}
\begin{center}
 \fontsize{12pt}{20pt}\selectfont{Christopher Couzens\textsuperscript{\orcidlink{0000-0001-9659-8550}}\footnotetext{\href{mailito:christopher.couzens@maths.ox.ac.uk}{christopher.couzens@maths.ox.ac.uk}}, Alice L\"uscher\textsuperscript{\orcidlink{0009-0001-8231-3080}}\footnotetext{\href{mailito:alice.luscher@maths.ox.ac.uk}{alice.luscher@maths.ox.ac.uk}} and James Sparks\textsuperscript{\orcidlink{0000-0003-3699-5225}} \footnotetext{\href{mailito:james.sparks@maths.ox.ac.uk}{james.sparks@maths.ox.ac.uk} }}

\end{center}
\vskip .2cm

 \vspace*{.5cm} 
 \fontsize{12pt}{14pt}\selectfont{{ Mathematical Institute, University of Oxford,\\ Andrew Wiles Building, Radcliffe Observatory Quarter,\\ Woodstock Road, Oxford, OX2 6GG, U.K.}\\}

 {\tt {}}

\vspace*{0.8cm}
\end{center}

 \renewcommand{\thefootnote}{\arabic{footnote}}
 
\begin{center} {\bf Abstract } 
\end{center}
We construct equivariantly closed polyforms for the fluxes and the action of (massive) type IIA and type IIB supergravity directly in ten dimensions. We illustrate applications of our formalism to two distinct classes of solutions in massive type IIA. Our analysis straightforwardly reproduces and vastly generalizes all known results for on-shell actions in these classes, with no need for a consistent truncation. 
Where known, we match the corresponding field theory partition functions at leading order, and our results extend easily to other solutions of type II supergravity.

\noindent 

\newpage
%%%%%%%%%%%%%%%%%%%%%%%%%%%%%%%%%%%%%%%%%%%
%%%           TITLE ENDS HERE
%%%%%%%%%%%%%%%%%%%%%%%%%%%%%%%%%%%%%%%%%%%
\vspace{-3mm}
\tableofcontents
\printindex

%%%%%%%%%%%%%%%%%%%%%%%%%%%%%%%%%%%%%%%%%%%
%%%           Intro
%%%%%%%%%%%%%%%%%%%%%%%%%%%%%%%%%%%%%%%%%%%
\newpage
\section{Introduction}

There are two main classes of solutions of interest when considering supersymmetric backgrounds relevant for holography and AdS/CFT. The first class studies supersymmetric AdS solutions embedded in ten- and eleven-dimensional supergravity, i.e.\ backgrounds of the form $\text{AdS}_p\times X_{D-p}$. Via the AdS/CFT correspondence these are dual to $(p-1)$-dimensional CFTs on a Minkowski background. The details of the CFT, for example the symmetries and field content, are encoded in the choice of compact internal space $X_{D-p}$ and one can learn about the supersymmetric protected observables such as anomalies, free energy and conformal dimensions of BPS operators by studying this geometry. There is a rich history of classifying the most general supersymmetry-preserving AdS solutions using G-structure techniques \cite{Gauntlett:2002sc,Gauntlett:2003cy,Gauntlett:2002fz}, and more recently generalized complex geometry \cite{Grana:2004bg,Tomasiello:2011eb}. 

The second class of solutions involves replacing AdS with a space $M_p$ which is asymptotically locally AdS. One now fibres $X_{D-p}$ over $M_p$ so that the spacetime takes the form  $M_{p}\ltimes X_{D-p}$, where $X_{D-p}$ may also be warped. This is then interpreted as placing the CFT on the non-trivial background $\partial M_p$ and tells us about deformations of the CFT. This allows us, for example, to probe how the observables of the CFT depend on the background, and we can compute protected observables such as indices and twisted partition functions. Typically one works in a $p$-dimensional gauged supergravity theory which is a consistent truncation of the $D$-dimensional theory. 

While consistent truncations provide an invaluable bridge between higher-dimensional string theory and lower-dimensional gauged supergravity, they are only available for special internal manifolds and restricted field content. Indeed, for some of the examples we will present, such consistent truncations are not currently known. Furthermore, many questions can only be answered when studying the higher-dimensional parent theories. For example, studying probe branes in a black hole background requires analysing the embedding of the branes in the full parent theory, and not merely within the consistent truncation of which the black hole is a solution  \cite{Aharony:2021zkr,Chen:2023lzq,Cabo-Bizet:2023ejm,BenettiGenolini:2026hmz,Suh:2026ikt}. Moreover, questions such as studying resolutions of punctured Riemann surfaces require a higher-dimensional embedding (or coupling the gauged supergravity to non-Abelian vector multiplets with no known uplifts \cite{DeLuca:2018zbi,Bobev:2018sgr}). For these, and many other reasons, a higher-dimensional description, involving both the external and internal spacetime, is thus desirable. This, however, is no easy task, as one needs to solve Einstein's equations in ten and eleven dimensions. Even with supersymmetry one still typically has non-linear partial differential conditions to solve.

Equivariant localization in supergravity, first developed in \cite{BenettiGenolini:2023kxp}, is a new method allowing for the computation of supergravity observables without the need to solve the equations of motion explicitly. This has been applied to a variety of setups, either in gauged supergravity or for the internal spaces of classes of AdS solutions \cite{BenettiGenolini:2023kxp,BenettiGenolini:2023yfe,BenettiGenolini:2023ndb,BenettiGenolini:2024kyy,BenettiGenolini:2024xeo,Cassani:2024kjn,Colombo:2023fhu,Couzens:2024vbn,Martelli:2023oqk,Suh:2024asy,Hristov:2024cgj,BenettiGenolini:2024hyd,BenettiGenolini:2024lbj,Couzens:2025ghx,Colombo:2025ihp,Cassia:2025jkr, BenettiGenolini:2025icr,Couzens:2025nxw, Colombo:2025yqy, Park:2025fon,Hristov:2025ygn,Hristov:2026tde,BenettiGenolini:2026qdm,Gaar:2026nqq,BenettiGenolini:2026hmz,Galli:2026lsl,Cassani:2026teb}. To apply the method one requires the existence of a symmetry, with the Berline--Vergne--Atiyah--Bott (BV--AB) theorem stating that the integral receives contributions only from the fixed points of the symmetry. Typically, supersymmetric solutions possess such a symmetry, dual to the R-symmetry, with the vector field generating the symmetry constructed from spinor bilinears. One then constructs equivariantly closed polyforms with respect to this symmetry with the top-form the observable of interest. 

We will apply this formalism directly to (massive) type IIA and type IIB supergravity. This allows us to simultaneously address both cases outlined above. For supersymmetric solutions preserving a minimal amount of supersymmetry, the ten-dimensional manifold generically has a Killing vector \cite{Tomasiello:2011eb}. We use a subset of the necessary and sufficient conditions for such a supersymmetric solution to exist \cite{Tomasiello:2011eb,Legramandi:2018qkr} to compute equivariantly closed polyforms for the Page fluxes and the on-shell action. This allows us to appropriately compute the quantization of fluxes for any supersymmetric supergravity background and to then compute the on-shell action. We find that there is a very universal structure for these polyforms.
Our results are applicable to whatever topologies and fluxes one wishes to consider. We will apply this formalism to geometries with asymptotically AdS spaces; however, it is applicable more broadly, including to asymptotically flat black holes and even non-conformal branes. We hope to present results in these directions in the near future.

The rest of the paper is organized as follows. In section \ref{sec:10dpoly}, we first review the supergravity backgrounds of \cite{Tomasiello:2011eb}, and further exploit their structure to build ten-dimensional equivariantly closed polyforms for the fluxes and on-shell actions in type II. 
The next two sections are dedicated to utilizing these polyforms for two classes of solutions in massive type IIA. In section~\ref{sec:6+4} we consider a hemisphere $HS^4$ fibration over a six-dimensional spacetime, giving rise to an extension of Romans supergravity, while section \ref{sec:4+6} focuses on $X_6$ fibrations over a four-dimensional spacetime where $X_6$ is the suspension of a five-dimensional Sasaki--Einstein manifold. In both sections, we derive a master formula computing the on-shell action for any topology, and further illustrate it for specific examples. Technical details are relegated to various appendices.

%%%%%%%%%%%%%%%%%%%%%%%%%%%%%%%%%%%%%%%%%%%
%%%           10d polyforms
%%%%%%%%%%%%%%%%%%%%%%%%%%%%%%%%%%%%%%%%%%%

\section{Ten-dimensional polyforms}\label{sec:10dpoly}

We are interested in computing the (partially) on-shell action of supersymmetric solutions of type IIA and type IIB supergravity. In order for the solutions to be well-defined string backgrounds we must also quantize the fluxes. Our goal in this section is to construct equivariantly closed polyforms for the Page fluxes before using these to construct a set of equivariantly closed polyforms for the partially on-shell action. A key tool in constructing these polyforms is the classification of supersymmetric solutions using generalized geometry \cite{Tomasiello:2011eb,Legramandi:2018qkr}. To keep the section self-contained we will first review the necessary conditions for a solution to preserve supersymmetry. 

%%%%%%%%%%%%%%%%%%%%%%%%%%%%%%%%%%%%%%%%%%%
%%%           Review of susy in 10d
%%%%%%%%%%%%%%%%%%%%%%%%%%%%%%%%%%%%%%%%%%%
\subsection{Preserving supersymmetry in ten dimensions}

In the following we will consider the two type II theories concurrently, and in order to do so we must introduce a little notation. Let $g$ be the metric, $\phi$ the dilaton, and $H=\dd B$ the NS-NS three-form. In addition, using the democratic approach, we consider the polyform of all RR fluxes given by
\begin{equation}
    \Fm=\sum_{k} \Fm_k\, ,
\end{equation}
where $k$ runs over all even numbers between 0 to 10 for type IIA and over all odd numbers in this range for type IIB. This democratic formulation doubles up the fluxes, where one must then impose the self-duality constraint
\begin{equation}\label{eq:selfdual}
    \Fm=\star\mskip2mu \lambda(\Fm)\, ,
\end{equation}
where $\lambda$ acts on a $k$-form as $ \lambda(\omega_k)\equiv (-1)^{\left\lfloor\tfrac{k}{2}\right\rfloor}\omega_k$.
Furthermore, we denote by $\dd_H$ the twisted differential operator defined by
\begin{equation}
    \dd_H=\dd-H\wedge\, .
\end{equation}
In the democratic formulation the Maxwell equations for the RR fluxes simply become
\begin{equation}\label{eq:FEOM}
    \dd_H \Fm=0\, .
\end{equation}

Supersymmetric solutions of type IIA and type IIB supergravity admit two Majorana--Weyl Killing spinors $\epsilon_1$ and $\epsilon_2$, which in our conventions satisfy:
\begin{equation}
    \begin{cases}
        \Gamma_{11} \epsilon_1=\epsilon_1\, , \quad \Gamma_{11}\epsilon_2=-\epsilon_2 & \text{type IIA}\, ,\\
         \Gamma_{11} \epsilon_1=\epsilon_1\, ,\quad \Gamma_{11}\epsilon_2=\epsilon_2 & \text{type IIB}\, .
    \end{cases}
\end{equation}
Following \cite{Martucci:2011dn,Tomasiello:2011eb,Legramandi:2018qkr}  one constructs the vector bilinears\footnote{In comparison with \cite{Tomasiello:2011eb,Legramandi:2018qkr} we have rescaled the bilinear forms by a factor of $32$, more in line with the conventions in \cite{Martucci:2011dn}.  }
\begin{equation}\label{eq:Kidef}
K_\alpha=\overline{\epsilon_\alpha}\Gamma^{M}\epsilon_\alpha\partial_{M}\, ,
\end{equation}
which are null as follows by the $\epsilon_\alpha$ being Majorana--Weyl. It follows that the Killing spinors satisfy the projection conditions
\begin{equation}
    K_{\alpha}\cdot \epsilon_{\alpha}=0\, ,
\end{equation}
with the dot denoting Clifford multiplication. 
One then constructs the vector field $K$ and one-form $\widetilde{K}$ defined as
\begin{equation}
    K=\frac{1}{2}(K_1+K_2)\, ,\qquad \widetilde{K}=\frac{1}{2}(K_1-K_2)\, .
\end{equation}
Since the $K_i$ are null, it follows that $K$ is either null (if the two $K_i$ are parallel) or otherwise time-like: $K^2< 0$. One can show that supersymmetry implies
\begin{equation}\label{eq:KHeq}
    \dd \widetilde{K}=K\hook H\, .
\end{equation}

Consider now the mixed bispinor
\begin{equation}
    \widehat{\Phi}=32\mskip3mu \epsilon_1\otimes \overline{\epsilon_2}\, ,
\end{equation}
which satisfies $\widehat{\Phi}=\star\mskip2mu \lambda(\widehat{\Phi})$.
One should understand this as a polyform upon application of the Clifford map. 
If the two Killing spinors have the same chirality then $\widehat{\Phi}$ is a sum of odd forms only, whilst if the spinors have opposite chirality then it is a sum of even forms only; therefore
\begin{equation}
    \widehat{\Phi}=\, 
    \begin{cases}
\, \widehat{\Phi}_0+\widehat{\Phi}_2+\widehat{\Phi}_4+\widehat{\Phi}_6+\widehat{\Phi}_8+\widehat{\Phi}_{10} & \text{type IIA}\, ,\\
  \,   \widehat{\Phi}_1+\widehat{\Phi}_3+\widehat{\Phi}_5+\widehat{\Phi}_7+\widehat{\Phi}_9 & \text{type IIB}\, .
    \end{cases}
\end{equation}
It follows that $\widehat{\Phi}$ satisfies 
\begin{equation}\label{eq:KsintoPsi}
    (\widetilde{K}\wedge + \mskip2mu K\hook\mskip3mu)\mskip2mu\widehat{\Phi}=0\, .
\end{equation}
One can derive a set of necessary and sufficient conditions for the preservation of supersymmetry. In particular, one finds that $K$ is a Killing vector for the full solution, the Killing spinors are uncharged under it, $\mathcal{L}_K\epsilon_i=0$, and that one necessarily has
\begin{equation}\label{eq:susyMVP}
    \dd_H\big(\me^{-\phi}\mskip2mu\widehat{\Phi}\big)=-(\widetilde{K}{\wedge}+K\hook\mskip3mu)\,\Fm\, .
\end{equation}
There are further conditions that need to be satisfied, known as pairing constraints, and we refer the reader to \cite{Tomasiello:2011eb,Legramandi:2018qkr} for the full set of necessary and sufficient conditions; however, it turns out that equations \eqref{eq:KHeq}, \eqref{eq:KsintoPsi} and \eqref{eq:susyMVP} are sufficient for us to construct the equivariantly closed polyforms that we will need.

%%%%%%%%%%%%%%%%%%%%%%%%%%%%%%%%%%%%%%%%%%%
%%%           Polyforms
%%%%%%%%%%%%%%%%%%%%%%%%%%%%%%%%%%%%%%%%%%%
\subsection{Constructing the polyforms}

Having reviewed the necessary background material we will turn to constructing equivariantly closed polyforms, with respect to the Killing vector $K$, for both type II theories. The key observation is that the supersymmetry conditions outlined above are precisely sufficient to find a completion of all the Page fluxes and the on-shell action into equivariantly closed polyforms. This then allows us to appropriately quantize the fluxes and then compute the on-shell action.
The rest of this section outlines the computation of these polyforms, with a more detailed exposition presented in appendix \ref{app:polyforms}.

First recall that in the presence of non-trivial NS-NS flux the Maxwell charges are not conserved, as is evident from \eqref{eq:FEOM}. Instead, the conserved charge is the Page charge. As such we are interested in constructing equivariantly closed polyforms for the Page fluxes. Using the democratic formulation, the Page flux polyform $F$ takes the simple form
\begin{equation}\label{eq:Pagedef}
    F=\me^{-B}\wedge \Fm\, ,
\end{equation}
where $B$ is the potential for the NS-NS three-form. From \eqref{eq:FEOM} it follows that $F$ is closed and therefore each of the terms of the same form degree may serve as the top component of an equivariantly completed polyform. 

Since $K$ preserves all of the bosonic fields we have that $\mathcal{L}_K H=0$. { Consequently, as $H=\dd B$, we can pick a gauge for $B$, such} that in each patch the Lie derivative of $B$ along the flow generated by $K$ also vanishes, {i.e.\ that} $\mathcal{L}_{K}B=0$.\footnote{{The existence of this gauge is a local statement. Indeed, using \eqref{eq:KHeq} and Cartan's formula gives $\mathcal L_K B=\dd(K\hook B)+K\hook H=\dd(K\hook B+\widetilde K).$ To set this to zero by a gauge transformation $B\to B+\dd\Lambda$ one must solve $\dd\big(\mathcal L_K\Lambda+K\hook B+\widetilde K\big)=0$, which can always be done locally, but may be obstructed globally.}} In this gauge, and using \eqref{eq:KHeq}, one finds
\begin{equation}
    0=\mathcal{L}_K B= \dd(\widetilde{K}+K\hook B)\, ,
\end{equation}
and we may therefore locally introduce a function $\y$ such that 
\begin{equation}\label{eq:ydef}
    \widetilde{K}+K\hook B=\dd \y\, .
\end{equation}
Using \eqref{eq:susyMVP} and \eqref{eq:ydef} we find that $K$ contracted into the Page fluxes is
\begin{equation}
    K\hook F=-\dd \Big[\y F+\Phi\Big]\, ,
\end{equation}
where we defined the polyform
\begin{equation}\label{eq:Phidef}
    \Phi\equiv \me^{-\phi}\mskip2mu\me^{-B}\wedge \widehat{\Phi}\, .
\end{equation}
The term in the brackets is the first step in constructing the equivariantly closed polyforms for the Page fluxes. We can then compute $K$ contracted into this polyform and proceed further along the descent -- the reader interested in these details should consult appendix \ref{app:polyforms}. The final result is a polyform {composed} of polyforms for the Page fluxes:
\begin{equation}\label{fluxpolypolyform}
    \Psi^{F}=F+\sum_{m}^{}\Psi_{m}\, ,
\end{equation}
where $m$ runs from 1 to 5 in IIA and 1 to 4 in IIB, while
\begin{align}\label{fluxpolypolyformcomponents}
    \Psi_m &=\frac{(-1)^{m}}{m!}\Big(\y^{m} F+ m\y^{m-1}\Phi\Big )\, .
\end{align}
This is the central result of this paper and applies to the Page fluxes of both (massive) type IIA and type IIB supergravity under the single assumption of minimal supersymmetry.

The equivariantly closed polyform for the Page flux $F_k$ is extracted by taking the sum of the $k$-form of $F$, i.e. $F_k$, the $(k-2)$-form of $\Psi_1$, the $(k-4)$-form of $\Psi_2$ and so forth until one ends up with either a zero-form for type IIA or a one-form for type IIB. 
Explicitly, the polyforms for the first three Page fluxes in massive type IIA are 
\begin{align}\label{eq:IIAfluxpoly}
        \Psi^{F_0}&= F_0\, ,\\ \nn
        \Psi^{F_2}&= F_2-(\y F_0+\Phi_0)\, ,\\ \nn
        \Psi^{F_4}&=F_4-(\y F_2+\Phi_2)+\frac{1}{2}(\y^2 F_0+2 \y \Phi_0)\, ,
\end{align}
where we have used the short-hand $\Phi_k=\Phi\mskip2mu|_{k\text{-form}}$.
The polyforms for the higher-form fluxes, which we also use later, and the ones for type IIB have the same structure and are spelled out explicitly in appendix \ref{app:polyforms}.
One should then quantize the $k$-form Page flux as
\begin{equation}
    \frac{1}{(2\pi\ls)^{k-1}}\int_{\Sigma_k}F_k\in\mathbb{Z}\, .
\end{equation}
In particular, one can quantize the Romans mass and define 
\begin{align}
n_0=2\pi \ell_s F_0\, .
\end{align}
We emphasize that these polyforms are valid for any minimally supersymmetric massive type IIA background, and those in appendix \ref{app:polyforms} hold for any minimally supersymmetric type IIB background.
Not only will these polyforms be instrumental in quantizing the Page fluxes, they are also sufficient for computing the on-shell action to which we now turn.

%%%%%%%%%%%%%%%%%%%%%%%%%%%%%%%%%%%%%%%%%%%
%%%           Partially on-shell action
%%%%%%%%%%%%%%%%%%%%%%%%%%%%%%%%%%%%%%%%%%%
\subsection{Partially on-shell action}

The key observable that we wish to compute in this paper is the (partially) on-shell string frame action of type II supergravity. For both type II theories, using the democratic formulation one can write a pseudo action which takes the form
\begin{equation}\label{pseudoaction}
    S=\frac{1}{(2\pi)^7\ls^8}\int\dd^{10} x \sqrt{-g}\bigg[ \me^{-2\phi}\Big(R+4 |\dd\phi|^2 -\frac{1}{2}|H|^2\Big) -\frac{1}{4} \sum_{k}|\Fm_k|^2\bigg]\, ,
\end{equation}
where for a (real) $p$-form $\alpha$ we define $|\alpha|^2 \equiv\frac{1}{p!}\alpha_{\mu_1\cdots\mu_p}\alpha^{\mu_1\cdots\mu_p}$.
By doubling the number of fields we remove the Chern--Simons terms from the action, which are effectively hidden in the last term \cite{Bergshoeff:2001pv}. 
Notice  that it is the Maxwell fluxes that appear in \eqref{pseudoaction}, rather than the Page fluxes we have just computed equivariantly closed polyforms for. 

The action
\eqref{pseudoaction} is a pseudo action in the sense that we are summing over all the RR fluxes but have not imposed the self-duality constraint in \eqref{eq:selfdual}. 
Imposing this at the level of the action implies that the RR term \emph{vanishes}, much like the usual story in type IIB supergravity for the self-dual five-form $F_5$. On the other hand, the latter has been addressed recently in the holographic context in \cite{Kurlyand:2022vzv,Adhikari:2026rfb}, where it is shown that the correct on-shell action for $\text{AdS}_5\times M_5$ solutions in type IIB supergravity, supported by self-dual five-form flux, is given by splitting the flux into electric and magnetic components. 
Their result for the on-shell action is derived from the PST formalism \cite{Pasti:1996vs}, which requires an additional boundary term. Rather than attempting to carry out a similar analysis for type II (using for example\footnote{This 
can also be made supersymmetric, and we expect the polyforms we have constructed to generalize appropriately. We thank Dmitri Sorokin for discussions on this.}  \cite{Mkrtchyan:2022xrm}), we instead propose that the analogous result holds generally: namely one uses the result \eqref{pseudoaction}, but 
separates the RR contributions into electric and magnetic, only including half the terms in the sum. Effectively, the RR terms in \eqref{pseudoaction} cancel pairwise on-shell, and this prescription means one is choosing to include only half the terms. 
The inherent ambiguity in the signs in doing this has a physical interpretation: the different choices are related by a Legendre transform that interchanges between electric and magnetic ensembles \cite{Hawking:1995ap}. For the purposes of the current paper we take this as a conjectured prescription for the on-shell action, and show this agrees 
both with lower-dimensional supergravity results obtained with holographic renormalization (when there is a consistent truncation), and also with field theory results that otherwise have no known gravity dual. We find perfect agreement with all classes of known supergravity solutions and large $N$ field theory  results. 

Inserting the dilaton equation of motion and dropping a boundary term (see appendix~\ref{app:action}), the above
prescription for the action for type IIA takes the form 
\begin{equation}
     S^\mathrm{IIA}=\frac{1}{2(2\pi)^7\ls^8}\int_{M_{10}}\Big[s_1 F_0 F_{10}+s_2 F_2\wedge F_8+s_3 F_4\wedge F_6\Big]\, ,
\end{equation}
and for IIB
\begin{equation}
     S^\mathrm{IIB}=\frac{1}{2(2\pi)^7\ls^8}\int_{M_{10}}\Big[s_1 F_1 \wedge F_{9}+s_2 F_3\wedge F_7+s_3 F_5\wedge F_5\Big]\, ,
\end{equation}
where the $s_i$'s are the signs depending on the ensemble that we are working in, i.e.\ which fluxes we held fixed. For all $s_i$ being $+1$ we are in the ensemble where $F_0$, $F_2$ and $F_4$ define conserved magnetic charges which we hold fixed in IIA (or $F_1$, $F_3$, $F_5$ in IIB).\footnote{In particular we always have $s_1=+1$ for IIA and $s_3=+1$ for IIB.} Other ensembles are then related by the change of sign of the relevant term. 

Since the partially on-shell action is given by a sum over the wedge products  of equivariantly closed forms, we can simply compute the equivariantly closed polyform for the action. It is given by taking the wedge product of the equivariantly closed Page flux polyforms. For massive type IIA we find
\begin{equation}\label{eq:PsiI}
    \Psi= \Big[s_1 F_0 \Psi^{F_{10}}+s_2 \Psi^{F_2}\wedge \Psi^{F_8}+s_3 \Psi^{F_4}\wedge \Psi^{F_6}\Big]\, .
\end{equation}
We thus obtain a method for both quantizing the fluxes and computing the partially on-shell action. In the remainder of this paper we will exhibit the power of this formalism by computing the on-shell action of a variety of different classes of solution in type IIA. The IIB action can in principle be evaluated in a similar fashion. However, we postpone this to future work and just discuss the (massive) type IIA theory here.

In particular, we will focus on two classes of solutions: hemisphere $HS^4$ fibrations over $M_6$ and $S^6$ fibrations over $M_4$. For these two classes the signs in the action are respectively $\{s_1=1,s_2=1,s_3=+1\}$ and $\{s_1=1,s_2=1,s_3=-1\}$. The physical difference between the two classes is clear: in the first class $F_4$ is a magnetic flux which we hold fixed, whilst in the second $F_6$ defines a magnetic charge that we hold fixed. The change in ensemble precisely flips the sign $s_3$. For concreteness let us write the zero-form part of the action polyform for each of these
\begin{equation}
    \Psi^\mathrm{elec}_0=-\frac{2}{15}\y^3(F_0^2\y^2+5F_0\y\Phi_0+5\Phi_0^2)\,, \quad \Psi^\mathrm{mag}_0=\frac{1}{30}\y^3(F_0^2\y^2+5F_0\y\Phi_0+10\Phi_0^2)\,. 
\end{equation}
Similar expressions hold for the higher-form parts. Moreover, turning off the Romans mass these expressions simply reduce to
\begin{equation}
    \Psi^\mathrm{elec}_0=-\frac{2}{3}\y^3\Phi_0^2\,, \quad \Psi^\mathrm{mag}_0=\frac{1}{3}\y^3\Phi_0^2\,,
\end{equation}
which are then ``obviously" related by electro-magnetic duality. {Indeed, the two results are related by a Legendre transform, schematically $-2/3+1=1/3$.} Using the embedding of the massless theory into 11d supergravity as spelt out in \cite{Legramandi:2018qkr} these results should allow for the computation of both the on-shell action and flux quantization in M-theory.

As a final comment, in order to compare with field theory results we are really interested in computing the Euclidean on-shell action rather than the Lorentzian one. As such we need to perform a Wick rotation of the theory. This carries some subtleties in the properties of the spinors used to define the bilinears, where in particular there is no Majorana condition in ten-dimensional Euclidean space. One can follow the logic in \cite{Bergshoeff:2007cg,DHoker:2025nid} and complexify the theory working on a ``holomorphic" slice and then restricting the theory to Euclidean signature. The Killing vector $K$ is then taken to be complex in the Euclidean theory. Furthermore, we will use that this prescription multiplies the Lorentzian on-shell action by an overall $-\ii$ and to distinguish the two we will denote the Euclidean (partially) on-shell action as $I=-\ii S$. In summary, we will use localization to evaluate
\begin{equation}\label{eq:onshellI}
    I = \frac{-\ii}{2(2\pi)^7\ls^8}\int_{M_{10}}\Psi\,.
\end{equation}

%\subsection{Summary}

Before we move on to using our results, let us quickly summarize this section. 
Given any minimally supersymmetric type II background, the Page fluxes admit the universal equivariantly closed completion given in \eqref{fluxpolypolyform} and \eqref{fluxpolypolyformcomponents}. Using these one should quantize the fluxes over all compact cycles and use them to impose the homology relations of the spacetime under consideration.\footnote{This step of imposing all of the homology relations is not strictly necessary and typically follows from extremization of the action subject to the flux quantization conditions (this was first pointed out in \cite{Couzens:2025nxw}). The benefit of imposing all homology relations is that it typically simplifies the final extremization problem.  } One can then use this to compute the partially on-shell action \eqref{eq:onshellI} with the equivariantly closed polyform given in \eqref{eq:PsiI}. The action depends on a set of signs which are related to the ensemble in which one is computing.

%%%%%%%%%%~~~~~~~~~~~~~~~~~~~~~~~~~~~~~~~~~~~~~~~~~~~~~~~~~~~~~~~~~~~~~~~~~~
%%%%%                      10=6+4
%%%%%%%%%%~~~~~~~~~~~~~~~~~~~~~~~~~~~~~~~~~~~~~~~~~~~~~~~~~~~~~~~~~~~~~~~~~~

\section{10 = 6+4: Romans supergravity coupled to a vector}\label{sec:6+4}

As a first application of our polyforms we will consider solutions of Euclidean massive type IIA supergravity which are a four-dimensional hemisphere bundle\footnote{This is the only solution which admits an AdS$_6$ vacuum solution in (massive) type IIA \cite{Passias:2012vp}.} over a six-dimensional base $M_6$. When the bundle is taken to be trivial and $M_6$ is Euclidean AdS$_6$ this is the topology of the famous Brandh\"uber--Oz solution \cite{Brandhuber:1999np}. In this section we will consider $M_6$ to have a more general topology than just AdS$_6$ and allow for a general $HS^4$ fibration over $M_6$. We derive a master formula \eqref{Ifinal}, depending on the topology and global data of a putative solution, which can be written in terms of a function which we interpret as the prepotential of 6d U$(1)^2$ gauged supergravity. This generalizes the results of \cite{Couzens:2025ghx}, which considered the Abelian Romans theory.
We also expect to be able to derive our results directly from a six-dimensional $U(1)^2$ gauged supergravity theory, which couples the minimal Romans theory studied in \cite{Couzens:2025ghx} to an additional vector multiplet. This theory should be a consistent truncation of massive type IIA. However, note that we have no need for such a consistent truncation here -- we derive formulas by directly localizing in ten dimensions.

Consider the following embedding of the hemisphere in five dimensions 
\begin{align}HS^4\subset \C_1\oplus\C_2\oplus\R_{\geq 0}\, ,
\end{align}
with $|z_1|^2+|z_2|^2+x^2=1$, 
where $z_i\in\C_i$ are complex coordinates and $x\in \R_{\geq 0}$ is a polar coordinate on the hemisphere $HS^4$. The latter has boundary
$\partial HS^4=\{x=0\}\cong S^3$. One can define two two-dimensional hemispheres $HS^2_{i} = \{z_i=0\}$ within $HS^4$ which are trivial in relative homology $H_2(HS^4,S^3)$.  
We fibre $HS^4$ over $M_6$ using the $U(1)^2$ action on $\C_1\oplus\C_2$ via two complex line bundles $\mathcal{L}_i$ over $M_6$ with Chern classes $c_1(\mathcal{L}_i)\in H^2(M_6,\Z)$ for $i=1,2$. 

We define the projection $\pi: M_{10}\rightarrow M_6$ from the total space of the bundle to the base. One can then project the Killing vector $K$ down to the base $M_6$ so that
$\xi \equiv \pi_*K$ is a vector field on $M_6$. 
 Over a fixed point of $\xi$ on $M_6$, we decompose the action of $K$ on the normal space in $M_{10}$ as
\begin{equation}
K=\xi+\sum_{i=1}^{2}b_i \partial_{\varphi_i}\, \,, \quad \xi=\sum_a \varepsilon_a\del_{\psi_a}\,,
\end{equation}
where $\del_{\varphi_i}$ rotates $\mathcal{L}_i$ and we assume that $b_i\neq 0$ so that the fixed points of $K$ are then the fixed points of $\xi$ on $M_6$ at the pole of $HS^4$. The fixed point in $M_6$ can be part of a fixed submanifold of dimension $0$, $2$ or $4$, and the index $a$ correspondingly takes values  
 in $\{1,2,3\}$, $\{1,2\}$ or $\{1\}$ respectively, such that $\del_{\psi_a}$ rotates the line bundle $L_a$ in the normal space in $M_6$ over the fixed point set. 
The weights $\varepsilon_a$, $b_i$ at different fixed points are determined from the equivariant action of $K$ on $M_{10}$. 
In particular, they depend on the fixed point $p$ and to be precise we should write $b_i^p$ but will often suppress the index so as to not clutter formulae.
Since the Killing spinors are not charged under $K$ it follows that the weights at a fixed point must satisfy the constraint
\begin{equation}\label{eq:charge6d}
    \sum_a\sigma_a\varepsilon_a+\sum_{i=1}^2b_i=0\,,
\end{equation}
where $\sigma_a$ are signs depending on the properties of the spinor at the fixed point; see appendix~\ref{app:spinors}.

One can now analyse the contribution of the different fixed point sets to the on-shell action. The result of this analysis is the general master formula 
\begin{align}\label{Ifinal}
I  =  \frac{8}{27}F_{S^5}\Bigg\{&\sum_{\pt}\mp\frac{1}{\varepsilon_1\varepsilon_2\varepsilon_3}\mathcal{F}+ \sum_{\Sigma}\frac{1}{\varepsilon_1\varepsilon_2}\left[\sum_{i=1}^2p_i \partial_{b_i}\mathcal{F}-\sum_{a=1}^2\left(\frac{\ma_a}{\varepsilon_a}\right)\mathcal{F}\right] 
\nonumber\\
  + &\sum_{B_4}\mp\frac{1}{\varepsilon}\Bigg[\frac{1}{2}\sum_{i,j=1}^2 \langle p_i , p_j\rangle\partial^2_{b_i b_j}\mathcal{F} - \frac{1}{\varepsilon}\sum_{i=1}^2 
\langle p_i,\ma\rangle \partial_{b_i}\mathcal{F}+\frac{1}{\varepsilon^2}
\langle \ma,\ma \rangle\mathcal{F}\Bigg]\Bigg\}\, ,
\end{align}
where 
\begin{equation}
    \mathcal{F}\equiv\mathcal{F}(b_1,b_2)=(b_1b_2)^{3/2}
\end{equation}
is the prepotential on $HS^4$ and $F_{S^5}$ is the $S^5$ free energy
\begin{equation}\label{eq:FS5}
    F_{S^5}=-\frac{9 \sqrt{2}\pi N^{5/2}}{5 \sqrt{n_0}}\, .
\end{equation}
The parameters appearing are just the weights, $\varepsilon_a$, $b_i$ (which we emphasize depend on the fixed point and have a suppressed superscript) for the action of $K$ and the (associated) Chern classes $\ma_a$, $p_i$ for normal bundles inside $M_6$, or the $HS^4$ fibres, respectively. The $\langle ~, \,\rangle$ in the final line denotes 
the intersection pairing $H^2(B_4,\Z)\times H^2(B_4,\Z)\rightarrow \Z$. 
In the following, we will derive this result by
looking in turn at isolated fixed points $\pt$ of $\xi$ on $M_6$, fixed surfaces
$\Sigma=\Sigma_g$, and fixed four-manifolds $B_4$ giving rise to the various contributions in \eqref{Ifinal}. By way of exposition we present explicit applications of this formula for different topologies realizing each of the terms above. We take the Page fluxes 
(and their equivariant extensions) to represent relative classes in $H^*(M_{10},\partial)$ in what follows, where their restriction being zero on the D8/O8 boundary $\partial$ (an $S^3\subset HS^4$ bundle over $M_6$) is a boundary condition. We note this is satisfied for 
uplifted solutions to the Romans theory.

\subsection{Isolated fixed points}

Consider a fixed point $\pt\in M_6$ of $\xi$ and consider the copy of 
$HS^2_i$ lying above it in the fibre. 
On one hand the flux of $F_2$ through such a relative cycle is trivial, since $[F_2]\in H^2(M_{10},\partial)$, and on the other hand, it can be evaluated using equivariant localization. Combining these observations, and taking $\Phi_0$ to vanish on the boundary, immediately gives 
\begin{align}\label{F0yp}
0 = \int_{HS^2_i} F_2 =-\frac{2\pi}{b_i} \Psi^{F_2}_0|_\pt= -\frac{2\pi}{b_i}  (F_0 \mskip1mu \y + \Phi_0) |_\pt 
\quad \Rightarrow \quad F_0\mskip1mu \y |_\pt = -{\Phi}_0 |_\pt\, .
\end{align}
Next we consider the flux of $F_4$ through the hemisphere, again at the fixed point\footnote{Note that these hemispheres are homologous as classes in $H_4(M_{10},\partial)$ (taking  
$M_6$ to be connected) so that $N$ is independent of $\pt$.} $\pt$
\begin{align}
N \equiv \frac{1}{(2\pi \ell_s)^3}\int_{HS^4}F_4 = \frac{1}{(2\pi \ell_s)^3}\frac{(2\pi)^2}{b_1b_2}(F_0\mskip1mu \tfrac{1}{2}\y^2 + 
\Phi_0\mskip1mu \y)\mid_\pt\, ,
\end{align}
and solving using \eqref{F0yp} gives
\begin{align}\label{ypm}
\y |_\pt = \pm \frac{2\sqrt{2}\pi \ii \ell_s^2}{\sqrt{n_0}} \sqrt{b_1 b_2}\mskip2mu N^{1/2}\, .
\end{align}

This is enough to turn to the evaluation of the action.
If $\pt\in M_6$ is an isolated fixed point, with weights $\varepsilon_a$, $a=1,2,3$,
for $K$ on the normal space, then we can immediately localize
\begin{align}\label{Ipv1}
I_{\pt} = \frac{\ii}{2(2\pi)^7\ls^8}\frac{(2\pi)^5}{\varepsilon_1 \varepsilon_2\varepsilon_3b_1b_2} \Psi_0^\mathrm{elec} |_\pt  = \mp \frac{8 \sqrt{2}\pi}{15\sqrt{n_0}}\frac{(b_1b_2)^{3/2}}{\varepsilon_1\varepsilon_2\varepsilon_3} N^{5/2}\, .
\end{align}
Here the sign is that in \eqref{ypm}.
Note that scaling 
all five weights by the same constant leaves both \eqref{eq:charge6d} and the action  contribution \eqref{Ipv1} invariant, as it should, so this choice of normalization is irrelevant. 

To obtain the on-shell free energy for EAdS$_6$, with boundary five-sphere, one should extremize the above subject to the sum of the weights vanishing. Doing this one obtains \eqref{eq:FS5}.
This is the free energy of the dual five-dimensional SCFT on $S^5$.
We may finally rewrite the off-shell action result for an arbitrary isolated fixed point in the form
\begin{align}\label{Ip}
I_{\pt}  = \mp\frac{8}{27}F_{S^5} \frac{\mathcal{F}(b_1^\pt,b_2^\pt)}{\varepsilon_1^\pt\varepsilon_2^\pt\varepsilon_3^\pt}\, ,
\end{align}
where we re-establish the superscript $\pt$ as a reminder that the weights in general depend on the fixed point.

%%%%%%%%%%~~~~~~~~~~~~~~~~~~~~~~~~~~~~~~~~~~~~~~~~~~~~~~~~~~~~~~~~~~~~~~~~~~
%%%%%                       Spindle
%%%%%%%%%%~~~~~~~~~~~~~~~~~~~~~~~~~~~~~~~~~~~~~~~~~~~~~~~~~~~~~~~~~~~~~~~~~~

\paragraph{Spindle}

As an example, we can consider the case $M_6=\mathbb{R}_4\times\Sigma$, where $HS^4$ is fibred over the spindle $\Sigma\equiv\mathbb{WCP}^1_{[n_+,n_-]}$.
The hemisphere directions $\R^2_1\oplus\R^2_2=\C_1\oplus\C_2$ are fibred as $\mathcal{O}(-p_1)\oplus\mathcal{O}(-p_2)\rightarrow\Sigma$, where the twist and anti-twist are $p_1+p_2=n_++\sigma n_-$ with $\sigma=\pm 1$. Here the poles of $\Sigma$ are denoted with $\pm$ (not to be confused with the above sign) and are fixed under the action of $\xi$. 
The weights at the two fixed points 
are then $(\varepsilon_1,\varepsilon_2,\mp \varepsilon/n_{\pm},b_1^{\pm},b_2^{\pm})$, with $\varepsilon$ the weight of $\xi$ in the spindle direction.
The twisting precisely means that 
 \begin{align}
b_i^+-b_i^- = \frac{p_i}{n_+n_-}\varepsilon\, ,
 \end{align}
 while taking  $\sigma_a=1$,
the charge condition \eqref{eq:charge6d} gives
\begin{align}\label{spindlecharge}
\varepsilon_1+\varepsilon_2 + b_1^+ + b_2^+ -  \frac{\varepsilon}{n_+} & = 0 \, ,\quad
\varepsilon_1+\varepsilon_2 + b_1^- + b_2^- + \sigma \frac{\varepsilon}{n_-}  = 0 \, .
\end{align}
Introducing appropriate orbifold factors in \eqref{Ip} and recalling  $\varepsilon_3=\mp \varepsilon/n_\pm$, the total action then reads
 \begin{align}\label{spindleac}
  I =  -\frac{8}{27}F_{S^5}\frac{(b_1^+b_2^+)^{3/2}-\sigma (b_1^-b_2^-)^{3/2}}{\varepsilon_1\varepsilon_2\varepsilon}\, .
 \end{align}
In particular, the EAdS$_4$ case has 
$\varepsilon_1=\varepsilon_2=-\frac{1}{2}$ and the action matches the spindle solution in \cite{Couzens:2024vbn}.\footnote{We also recognize \eqref{spindlecharge} 
as precisely equation (4.23) in 
\cite{Couzens:2024vbn} correcting 
the appearance of the twist/anti-twist choice $\sigma$ at the $-$ pole.} 
This can be viewed 
 as an EAdS$_4\times X_6$ solution with the internal space $X_6\equiv HS^4\ltimes\Sigma$, 
 or as a spindle solution to 
 Romans theory coupled to an additional 
 flavour vector multiplet (which 
 allows $p_1\neq p_2$).

 \paragraph{Squashed sphere}

The above analysis allows us to similarly recover the squashed sphere $S^3_b\times S^2_{\varepsilon}$ field theory
result of \cite{Hosseini:2021mnn}, with the squashing parameter 
 \begin{equation}\label{squashing}
   b\equiv \sqrt{\frac{\varepsilon_1}{\varepsilon_2}}\,,\quad Q^2 \equiv \frac{(\varepsilon_1+\varepsilon_2)^2}{4\varepsilon_1\varepsilon_2}\, .
\end{equation}
We set $n_+=1=n_-$ so that we have a sphere and look at the twist case $\sigma=+1$.
The variables are related in the following way
\begin{align}
   \mathfrak{t}_i=p_i\,,  \quad  \Delta_i^\pm = -\frac{2(b_i^\pm \mp \frac{\varepsilon}{2}p_i)}{\varepsilon_1+\varepsilon_2}\, , \quad 
    \tilde\varepsilon = \frac{2\varepsilon}{\varepsilon_1+\varepsilon_2} \,.
\end{align}
Then $\mathfrak{t}_1+\mathfrak{t}_2=2$, $\Delta_1^\pm + \Delta_2^\pm = 2$
and $\Delta_i^+-\Delta_i^-=0$, which 
matches the field theory variables. We can drop the $\pm$ superscripts, and write $\Delta_i^\pm \equiv \Delta_i$.  
Note the rescaling of $\varepsilon$ is required to match the correct twisting condition, where $\Delta_i\pm \frac{\tilde\varepsilon}{2}p_i$ appear in the two blocks. 

%%%%%~~~~~~~~~~~~~~~~~~~~~~~~~~~~
%%%%%
%%%%%~~~~~~~~~~~~~~~~~~~~~~~~~~~~

\subsection{Fixed surfaces}

Now let $\Sigma=\Sigma_g\subset M_6$ be a connected, compact surface 
that is fixed under $\xi$. This has normal bundle $N_{\Sigma\hookrightarrow M_{6}}=L_1\oplus L_2$ 
in $M_6$, where we may write $L_a=\mathcal{O}(-\ma_a)$, $a=1,2$ for 
$\ma_a\in\Z$. Similarly, the hemisphere line bundles restrict to 
$\mathcal{L}_i |_{\Sigma} = \mathcal{O}(-p_i)$, $i=1,2$, for 
$p_i\in\Z$. The total normal bundle of $\Sigma\subset M_{10}$ is 
then $N_{\Sigma\hookrightarrow M_{10}} = L_1\oplus L_2\oplus\mathcal{L}_1\oplus\mathcal{L}_2$.
Equivalently,
\begin{equation}
    \ma_a\equiv-\int_\Sigma c_1(L_a)\,,\quad p_i\equiv-\int_\Sigma c_1(\mathcal{L}_i)\,.
\end{equation}
We denote the weights 
on $L_a$ by $\varepsilon_a$, $a=1,2$, and the weights on 
the disc directions in the hemisphere by $b_i$, $i=1,2$, as before. Note 
these depend on the particular connected two-dimensional fixed 
point set $\Sigma$, but we suppress this in the notation. The weights are related through the charge constraint \eqref{eq:charge6d}. The latter extends to surfaces to further impose that the sum of the first Chern classes 
of the relevant line bundles
(including the one on $\Sigma$, which is equal to its Euler characteristic $\chi_\Sigma$) vanishes, such that (see appendix \ref{app:spinors})
\begin{equation}\label{genCY}
    m_1+m_2+p_1+p_2=\chi_\Sigma\,.
\end{equation}

We start by quantizing the fluxes through the various cycles.
Integrating ${F}_2$ through the $HS^2_i$ fibres 
over a point on $\Sigma$ gives the same condition as for an isolated fixed point in \eqref{F0yp}. Likewise integrating 
${F}_4$ through the $HS^4$ over a point on $\Sigma$ gives again 
\eqref{ypm}. However, we 
now have additional four-cycles to consider, which are  $HS^2_i$ fibred over $\Sigma$, and we denote these by $C_i$. On one hand, the flux through these cycles is quantized and can be computed using localization
\begin{align}
N_i \equiv \frac{1}{(2\pi \ell_s)^3}\int_{C_i} F_4 =\frac{1}{(2\pi \ell_s)^3}\int_{\Sigma^*} \left[\frac{2\pi}{b_i}\Psi^{{F}_4}_2 +\frac{(2\pi)^2}{b_i^2}\Psi^{{F}_4}_0 p_i\right]\, ,
\end{align}
where $\Sigma^*$ is the fixed copy of 
 $\Sigma$ at the pole $*$ of the $HS^2_i$. 
Note that the flux of ${F}_2$ through $\Sigma$ enters the first term, as well as $\Phi_2$, using \eqref{eq:IIAfluxpoly}.
On the other hand, we have the homology relations $N_1=-p_2 N$, $N_2=-p_1N$ , as reviewed in appendix~\ref{app:homology}. 
Combining these observations one can isolate 
\begin{align}\label{Phi2sub}
\int_{\Sigma^*}\Phi_2 = \mp\frac{2\sqrt{2}\ii\sqrt{b_1b_2}\ell_s^2\pi}{\sqrt{n_0}}N^{1/2}\int_{\Sigma^*}{F}_2+4\pi^2\ell_s^3(b_1 p_2+b_2p_1)N\, .
    \end{align}
The localized action is
\begin{align}
I_\Sigma = \frac{\ii}{2(2\pi)^7\ls^8}\frac{(2\pi)^4}{\varepsilon_1\varepsilon_2b_1b_2}\int_{\Sigma^*}\left[\Psi_2 + 2\pi\Psi_0 \left(\sum_{i=1}^2\frac{p_i}{b_i}+\sum_{a=1}^2\frac{\ma_a}{\varepsilon_a}\right)\right]\, .
\end{align}
Inserting the previous flux relations gives
\begin{align}\label{ISigma}
I_\Sigma %&= \frac{8}{27} F_{S^5} \frac{(b_1b_2)^{3/2}}{\varepsilon_1\varepsilon_2}\left[\frac{3}{2}\sum_{i=1}^2\frac{p_i}{b_i}-\sum_{a=1}^2\frac{\ma_a}{\varepsilon_a}\right]\\ \nn
&= \frac{8}{27} F_{S^5} \frac{1}{\varepsilon_1\varepsilon_2}\left[\sum_{i=1}^2 p_i\partial_{b_i}\mathcal{F}(b_1,b_2)-\mathcal{F}(b_1,b_2)\sum_{a=1}^2\frac{\ma_a}{\varepsilon_a}\right]\, ,
\end{align}
where $F_{S^5}$, given in \eqref{eq:FS5}, is used to eliminate $N$ and recall that $\mathcal{F}=(b_1b_2)^{3/2}$. Note that, unlike in the fixed point case, the dependence on the choice of sign in \eqref{ypm} drops out of the surface contribution.

%%%%%%%%%%~~~~~~~~~~~~~~~~~~~~~~~~~~~~~~~~~~~~~~~~~~~~~~~~~~~~~~~~~~~~~~~~~~
%%%%%                       Fixed B4
%%%%%%%%%%~~~~~~~~~~~~~~~~~~~~~~~~~~~~~~~~~~~~~~~~~~~~~~~~~~~~~~~~~~~~~~~~~~
\paragraph{Black brane} 

Consider $M_6=\mathbb{R}_4\ltimes\Sigma_g$, with the $HS^4$ also fibred over $\Sigma_g$. This corresponds to an asymptotically AdS$_6$ black brane, with horizon the Riemann surface $\Sigma_g$, magnetic charges $p_1$ and $p_2$ and the normal bundle twisting parameters $m_1$, $m_2$. Such a setup was previously studied in \cite{Couzens:2025ghx} for $b_1=b_2$, i.e.\ the Romans theory. On the other hand, turning off the $m$'s and setting $\varepsilon_1=\varepsilon_2=-\tfrac{1}{2}$, this is the EAdS$_4\times X_6$ solution with $X_6=HS^4\ltimes \Sigma_g$ studied in \cite{Couzens:2024vbn}. 
The charge constraints \eqref{eq:charge6d} and \eqref{genCY} (with $\sigma_i=1$) read
\begin{equation}
    \varepsilon_1+\varepsilon_2+b_1+b_2=0\,, \quad 
    m_1+m_2+p_1+p_2=\chi(\Sigma_g)\,.
\end{equation}
Note that for $m_1=m_2=0$, the latter is simply the Calabi--Yau condition for $X_6$. 

Taking the two weights $b_i$ equal and redefining the weights $\epsilon_i$ as
\begin{equation}
    \epsilon_i\equiv -2 w_i \frac{b_1}{w_1+w_2}
\end{equation}
trivializes the weight constraint and the on-shell action becomes
\begin{equation}
    I=\frac{F_{S^5}}{27 w_1^2 w_2^2}(w_1+w_2)^2\Big[3 w_1 w_2 \chi(\Sigma_g)+m_1 w_2(w_2-2 w_1)+m_2 w_1(w_1-2 w_2)\Big]\,,
\end{equation}
which agrees with the results in \cite{Couzens:2025ghx} -- see section (5.1) thereof.

%%%%%%%%%%~~~~~~~~~~~~~~~~~~~~~~~~~~~~~~~~~~~~~~~~~~~~~~~~~~~~~~~~~~~~~~~~~~
%%%%%                       Fixed B4
%%%%%%%%%%~~~~~~~~~~~~~~~~~~~~~~~~~~~~~~~~~~~~~~~~~~~~~~~~~~~~~~~~~~~~~~~~~~

\subsection{Fixed $B_4$}

Finally, let $B_4\subset M_6$ be a fixed connected, compact four-manifold 
under the action of $\xi$. This has normal bundle $N_{B_4\hookrightarrow M_6}=L$ a complex line bundle (on fixing an orientation), with Chern class $c_1(L)\in 
H^2(B_4,\Z)$. We pick a basis of two-cycles $\Gamma_\alpha\subset B_4$ which form a basis  
for the free part of $H_2(B_4,\Z)$, with Poincar\'e duals $\Psi_\alpha = \text{PD}[\Gamma_\alpha]\in H^2(B_4,\R)$. Given a closed\footnote{The zero-form parts of all RR equivariant forms are constant on a given connected component of the fixed point set, which implies that $\y$, $\Phi_0$ are constant. Then similarly one can deduce that $\Phi_2$ is closed from equivariant closure and $K$ being zero on a fixed point set.} two-form such as 
$\Phi_2 |_{B_4}$, we may write
\begin{align}\label{eq:HS4extra4cycles}
\phi^\alpha \equiv \int_{\Gamma_\alpha}\Phi_2 \quad \Rightarrow \quad 
[\Phi_2] = \sum_{\alpha=1}^{\text{b}_2}\hat{\phi}^\alpha\Psi_\alpha\in H^2(B_4,\R)\, ,
\end{align} 
where $\text{b}_2=\dim H_2(B_4,\R)$, and 
\begin{align}
\phi^\alpha = \int_{B_4}\Phi_2\wedge \Psi_\alpha = \sum_{\beta=1}^{\text{b}_2}Q_{\alpha\beta}\hat{\phi}^\beta\, ,
\end{align} 
where $Q_{\alpha\beta}\equiv \int_{B_4}\Psi_\alpha\wedge\Psi_\beta$ is the intersection matrix. This is an integer-valued unimodular symmetric matrix, with inverse 
$I=Q^{-1}$. Then, for example, one has
\begin{align}
\int_{B_4}\Phi_2 \wedge \Phi_2 = \sum_{\alpha,\beta=1}^{\text{b}_2}I_{\alpha\beta}
\phi^\alpha\phi^\beta \equiv \langle \phi, \phi\rangle \, .
\end{align}
We similarly introduce 
\begin{align}
k^\alpha \equiv \frac{1}{2\pi\ell_s}\int_{\Gamma_\alpha}F_2\, , \quad 
\ma^\alpha \equiv -\int_{\Gamma_\alpha}c_1(L)\, , \quad 
p_i^\alpha\equiv - \int_{\Gamma_\alpha}c_1(\mathcal{L}_i)\, .
\end{align}

Using our two-cycle basis $\Gamma_{\alpha}$ on $B_4$ we can define four-cycles $C^\alpha_i$ which are $HS^2_i$ fibred over $\Gamma_\alpha$. Note that these cycles satisfy the homology relations
 $N^\alpha_1 = -p^\alpha_2 N$,  $N^\alpha_2 = -p^\alpha_1 N$. 
The argument is the same as for fixed $\Sigma$, where now we simply apply this to 
$\Sigma=\Gamma_\alpha$. Then as before we can solve the analogue of \eqref{eq:HS4extra4cycles} to give
\begin{align}\label{Phi2subalpha}
\phi^\alpha = \mp\frac{4\sqrt{2}\ii\sqrt{b_1b_2}\ell_s^3\pi^2}{\sqrt{n_0}}k^\alpha N^{1/2}+4\pi^2\ell_s^3(b_1 p_2^\alpha+b_2p_1^\alpha)N\, .
    \end{align}
There is the copy of $B_4$ at the pole of $HS^4$, which we denote $B_4^*$. 
We then have the homology relation
\begin{align}
N^* \equiv \frac{1}{(2\pi\ell_s)^3}\int_{B_4^*}F_4 = \langle p_1,p_2\rangle N\, ,
\end{align}
which is proven similarly to the $S^4$ over $B_4$ case -- see appendix \ref{app:homology}. 

The localized action is then
\begin{align}
I_{B_4} & = \frac{\ii}{2(2\pi)^7\ell_s^8}\frac{(2\pi)^3}{\varepsilon\mskip2mu b_1b_2}\Bigg\{\int_{B_4^*}\left[\Psi_4 -2\pi\Psi_2\wedge \left(\frac{c_1(L)}{\varepsilon}+\sum_{i=1}^2\frac{c_1(\mathcal{L}_i)}{b_i}\right)\right]\nonumber\\
& \quad +(2\pi)^2\Psi_0\left(\frac{\langle p_1,p_1\rangle}{b_1^2}+\frac{\langle p_1,p_2\rangle}{b_1 b_2}+\frac{\langle p_2,p_2\rangle}{b_2^2}+\frac{\langle \ma,\ma\rangle}{\varepsilon^2}+\sum_{i=1}^2\frac{\langle \ma,p_i\rangle}{\varepsilon\mskip1mu b_i}\right)\Bigg\}\, .
\end{align}
After substituting in all the forms and the above formulae, we obtain the simple result
\begin{align}\label{IB4}
I_{B_4} & = \mp \frac{1}{\varepsilon}\Bigg[\frac{1}{2}\sum_{i,j=1}^2 \langle p_i , p_j\rangle\partial^2_{b_i b_j}\mathcal{F}(b_1,b_2) - \frac{1}{\varepsilon}\sum_{i=1}^2 
\langle p_i,\ma\rangle \partial_{b_i}\mathcal{F}(b_1,b_2)\nonumber\\
& \quad +\frac{1}{\varepsilon^2}
\langle \ma,\ma \rangle\mathcal{F}(b_1,b_2) \Bigg]\frac{8}{27} F_{S^5} \pm \frac{2\sqrt{2}\pi}{3\varepsilon}\sqrt{b_1b_2}\langle k, k \rangle\frac{N^{3/2}}{n_0^{3/2}} \, .
\end{align}
Note that the $F_2$ flux term $k^\alpha$ 
appears as an $N^{3/2}$ correction, and so is not seen at leading order in the 
large $N$ limit (assuming that the $F_2$ flux is order 1 at large $N$). It would be interesting to investigate such terms in more detail. {Such corrections are common in geometries where one includes these types of fluxes (see e.g.\ \cite{Honda:2014npa}); however, there may also be other terms contributing at this order, such as higher derivative corrections.} 

%%%%%%%%%%~~~~~~~~~~~~~~~~~~~~~~~~~~~~~~~~~~~~~~~~~~~~~~~~~~~~~~~~~~~~~~~~~~
%%%%%                       Fixed B4 Sg x Sg
%%%%%%%%%%~~~~~~~~~~~~~~~~~~~~~~~~~~~~~~~~~~~~~~~~~~~~~~~~~~~~~~~~~~~~~~~~~~

\paragraph{Product of Riemann surfaces}

As an example of the above formula consider $B_4$ to be the product of two Riemann surfaces: $B_4=\Sigma_{g_1}\times \Sigma_{g_2}$. For such a $B_4$ there are two two-cycles, represented by copies of the Riemann surface at a point on the other one. As such the intersection matrix takes the simple form
\begin{equation}
    I_{\alpha\beta}=\begin{pmatrix}
        0 &1\\
        1&0
    \end{pmatrix}_{\alpha\beta}\, .
\end{equation}
Then the bracket gives
\begin{equation}
    \langle p_i,p_j\rangle = p_i^{(1)}p_j^{(2)}+p_i^{(2)}p_j^{(1)}\,.
\end{equation}
The condition for the spinors to be well-defined, \eqref{eq:4df.p:c1}, implies that the line bundles over the Riemann surfaces are constrained to satisfy
\begin{equation}\label{eq:RiemRiemchi}
    \pm \chi(\Sigma_{g_\alpha})= \sigma_1 p_1^{\alpha}+\sigma_2 p_2^\alpha+\sigma_3 m^{\alpha}\, .
\end{equation}
One can now use this to expand out each of the terms in \eqref{IB4} and use \eqref{eq:RiemRiemchi} to eliminate one of the first Chern classes in favour of the Euler characteristic to get a more explicit, and substantially longer, result. 

 Setting $m=0$ so that 
$M_6=\Sigma_{g_1}\times\Sigma_{g_2}\times \R^2$ is simply a product, using \eqref{IB4} and taking the upper sign with $\sigma_1=\sigma_2=+1$ in \eqref{eq:RiemRiemchi} we find
\begin{align}\label{Sigma1Sigma2result}
I = \frac{2\sqrt{2}\pi}{5\sqrt{n_0}\mskip2mu \varepsilon}\frac{\left[b_2(b_2p_1^{(2)}+3b_1 p_2^{(2)})p_1^{(1)}+b_1(b_1p_2^{(2)}+3b_2p_1^{(2)})p_2^{(1)}\right]}{(b_1b_2)^{1/2}} N^{5/2}\, .
\end{align}
We may compare this to the large $N$ field theory result 
(3.105) for $\log Z$ derived in \cite{Hosseini:2018uzp,Suh:2018tul,Hosseini:2018usu,Suh:2018szn} for the dual $\text{USp}(2N)$ Seiberg theory on $\Sigma_{g_1}\times \Sigma_{g_2}\times S^1=\partial M_6$. Identifying our variables with theirs via $p_i^{(1)}=\mathfrak{s}_i$, $p_i^{(2)}=\mathfrak{t}_i$ and $b_i=\Delta_i$, our R-charge constraint 
$b_1+b_2=-\varepsilon$ matches their $\Delta_1+\Delta_2=2\pi$ provided we normalize the weight $\varepsilon=-2\pi$ 
on the $\R^2$ factor (with the Chern numbers satisfying \eqref{eq:RiemRiemchi}). The result \eqref{Sigma1Sigma2result}
then perfectly matches the field theory result for $-\log Z$. 

%%%%%%%%%%~~~~~~~~~~~~~~~~~~~~~~~~~~~~~~~~~~~~~~~~~~~~~~~~~~~~~~~~~~~~~~~~~~
%%%%%                       Fixed B4
%%%%%%%%%%~~~~~~~~~~~~~~~~~~~~~~~~~~~~~~~~~~~~~~~~~~~~~~~~~~~~~~~~~~~~~~~~~~

\section{10 = 4 + 6: D2-D8 Chern-Simons theories}\label{sec:4+6}

We now turn to a second application of our results, where the geometry is a fibration of the suspension $X_6=\mathcal{S}\mathrm{SE}_5$ over a four-dimensional base $M_4$ with $\mathrm{SE}_5$ a five-dimensional Sasaki--Einstein manifold. When $M_4$ is Euclidean AdS$_4$ this describes 3d $\mathcal{N}=2$ SCFTs arising from D2-branes probing the suspension of the Sasaki--Einstein manifold in the presence of D8-branes \cite{Fluder:2015eoa, Tomasiello:2010zz}. From the field theory perspective, the choice of SE$_5$ determines the matter content of the theory, while the choice of the four-dimensional spacetime means placing the dual theory on the background $\del M_4$.  

In this section we will focus on the $S^5$ case, and relegate performing the explicit computations for different Sasaki--Einstein manifolds ($T^{1,1}$ in particular) to appendix \ref{app:T11}. The resultant master formula for each Sasaki--Einstein space depends only on the definition of a prepotential given by the off-shell Sasakian volume to the power $-2/3$. For the moment we will keep the discussion general before specializing to the $S^5$ case. 

Let us first restrict to the toric Sasaki--Einstein case which allows us to introduce a basis for the U$(1)^3$ action given by $\del_{\varphi_i}$, $i=1,2,3$. For each of these directions we may introduce a complex line bundle $\mathcal{L}_i$ with which we can fibre over the base $M_4$. Let us introduce the projection $\pi:M_{10}\rightarrow M_4$ from the bundle to the base. Projecting the vector field $K$ to the base we define $\xi\equiv\pi_* K$ which is a vector field on $M_4$. We may then decompose the ten-dimensional Killing vector as:
\begin{equation}
K=\xi+\sum_{i=1}^{3}b_i \partial_{\varphi_i}\, ,\quad \xi=\sum_{a}\varepsilon_a\partial_{\psi_a}\,,
\end{equation}
where $a$ takes values $\{1,2\}$ for isolated fixed points in $M_4$ and $\{1\}$ for bolts in $M_4$.
As in the previous section it follows that the fixed points of $K$ are the fixed points of $\xi$ in $M_4$ at the two poles of the suspension. 

Following appendix \ref{app:spinors}, the fact that the Killing spinor is uncharged under $K$ fixes the weights at the fixed point to satisfy\footnote{We fix all of the signs for the weights on the $S^5$ to be $1$ at each fixed point of $M_4$.}
\begin{equation}\label{eq:charge4+6}
\sum_{a}\sigma_a\varepsilon_a+\sum_{i=1}^{3}b_i=0\, ,
\end{equation}
where $\sigma_a$ are signs which are fixed at each fixed point. 
Furthermore, when considering the weights of the suspension the product of the weights (independent of the number of weights used to compute a cycle) picks up an overall sign between the north and south pole. We will take the sign to be positive at the north pole (when computing using toric geometry) and negative at the south pole. Note that the weights in the external spacetime are the same at the north and south pole. 

For this class of theory it follows that the partially on-shell action is given by the following master formula: 
\begin{align}\label{Ifinal4d}
I  = \frac{9}{4}F_{S^3}\Bigg\{\sum_{\pt}\frac{1}{\varepsilon_1\varepsilon_2}\mathcal{F}+\sum_{\Sigma}\frac{1}{\varepsilon}\left[\sum_{i=1}^3p_i \partial_{b_i}\mathcal{F}-\left(\frac{\ma}{\varepsilon}\right)\mathcal{F}\right] 
\Bigg\}\, ,
\end{align}
where the prepotential is a homogeneous degree two function of the weights $b_i$, $i=1,2,3$ determined by the choice of SE$_5$ in terms of the Sasakian volume. For the five-sphere the prepotential is
\begin{equation}\label{prepotential}
    \mathcal{F}_{S^5}=(b_1b_2b_3)^{2/3}\,,
\end{equation} and $F_{S^3}$ is the free energy of the CFT dual to the $S^5$ suspension placed on $S^3$
\begin{equation}\label{SosS5susp}
    F_{S^3}=-\frac{2^{1/3}\pi}{5\times 3^{5/6}}(3\ii+\sqrt{3}) n_0^{1/3}N^{5/3}\,.
\end{equation}
Note in particular that $F_{S^3}$ is complex and it is a non-trivial result that our formalism recovers both the real and imaginary part. 

In the remainder of this section we prove the master formula \eqref{Ifinal4d} for $X_6=\mathcal{S}S^5=S^6$ and then generalize the analysis to other SE$_5$ suspensions in appendix \ref{app:T11}. The logic is similar to the previous section. We first quantize the fluxes and impose the homology relations before substituting this into the action and extremizing over the free parameters. As in the previous section we first recover the contribution of fixed points before turning to the fixed surface contributions. We provide examples to further illustrate both types of contributions, which recover and extend known results in the literature.

%%%%%%%%%%~~~~~~~~~~~~~~~~~~~~~~~~~~~~~~~~~~~~~~~~~~~~~~~~~~~~~~~~~~~~~~~~~~
%%%%%                      Isolated f.p.
%%%%%%%%%%~~~~~~~~~~~~~~~~~~~~~~~~~~~~~~~~~~~~~~~~~~~~~~~~~~~~~~~~~~~~~~~~~~
\subsection{Isolated fixed points}

Let us first quantize the various fluxes through the various cycles. There are three $S^2_i\subset S^6$ and similarly three $S^4_i\subset S^6$. These are all invariant under $K$ and trivial in homology, such that using localization 
\begin{align}
    &0=\int_{S^2_i} F^2=\frac{2\pi}{b_i}\left[(F_0 \y_N +\Phi_0^N)-(F_0 \y_S +\Phi_0^S)\right]\,, \\ \nn
    &0=\int_{S^4_i} F^4=\frac{(2\pi)^2}{b_jb_k}\frac{1}{2}\left[(F_0 \y_N^2 +2\y_N\Phi_0^N)-(F_0 \y_S^2 +2\y_S\Phi_0^S)\right]\,,
\end{align}
from which we obtain
\begin{equation}
    \Phi_0^N=-\Phi_0^S=-\frac{1}{2}F_0(\y_N-\y_S)\,.
\end{equation}
We now turn to the quantization of $F_6$ through $X_6$, which counts the number of D2-branes
\begin{align}
    N&=\frac{1}{(2\pi\ls)^5}\frac{(2\pi)^3}{b_1b_2b_3}\frac{1}{3!}\left[(F_0 \y_N^3 +3\y_N^2\Phi_0^N)-(F_0 \y_S^3 +3\y_S^2\Phi_0^S)\right] \\ \nn 
    &=\frac{1}{b_1b_2b_3}\frac{n_0}{96\ls^6\pi^3}(\y_N-\y_S)^3\,.
\end{align}
Inserting our previous result for $\Phi_0^{N,S}$, we can isolate an expression for $\y_N$. The flux being cubic in $\y_N$, it admits three roots
\begin{equation}\label{YNS5susp}
   % \y_N=-2^{2/3}(1-\ii\sqrt{3})\pi\ls^2\left(3b_1b_2b_3n_0^{-1}N\right)^{1/3} +\y_S\,.
   \y_N=-2 ^{5/3}\me^{2\pi \ii n_N/3}\pi \ls^2(3 b_1 b_2 b_3 n_0^{-1}N)^{1/3}+\y_S\, ,
\end{equation}
where $n_N=\{0,1,2\}$ and labels the three distinct roots.\footnote{The root with $n_N=0$ leads to a purely imaginary action and should be ignored.}

We have now all the ingredients to evaluate the action. Localization gives 
\begin{equation}
    I_\pt=\frac{\ii}{2(2\pi)^7\ls^8}\frac{(2\pi)^5}{\varepsilon_1\varepsilon_2b_1b_2b_3}(\Psi^\mathrm{mag}_0|_N-\Psi^\mathrm{mag}_0|_S)\,,
\end{equation}
where we insert the expressions for $\Phi_0^{N,S}$ and $\y_N$ obtained above. The explicit expression is not particularly enlightening as it stands, however extremizing over the remaining free parameter $\y_S$ we find that the critical point fixes
\begin{equation}\label{eq:extYS}
    \y_S=-\y_N\,.
\end{equation}
Inserting this back into the action gives the extremized result 
\begin{equation}\label{S5suspfixedpt}
I_{\pt}=(-1)^{n_N}\ii\me^{\ii \pi n_N/3}\frac{3^{5/3} \pi }{2^{2/3}5}n_0^{1/3}\frac{(b_1 b_2b_3)^{2/3}}{\varepsilon_1\varepsilon_2}N^{5/3}\, .
\end{equation}
Note that this is an off-shell result and there is still an extremization over the weights to perform. In order to obtain a result with positive real part one should take $n_N=1$ and the result is
\begin{equation}\label{eq:IS6finalexp}
   I_{\pt}=\frac{3^{7/6}(3-\ii \sqrt{3})\pi}{2^{5/3}\times 5}\frac{(b_1 b_2 b_3)^{2/3}}{\epsilon_1\epsilon_2}n_0^{1/3} N^{5/3}\, .
\end{equation}
On the other hand the choice $n_N=2$ switches the sign of the real part, playing the same role as the $\pm$ sign in the fixed point contribution \eqref{Ip} we obtained in the previous section. 

In the case of EAdS$_4$, there is a single fixed point and the extremization over the weights sets them to all be equal, such that \eqref{S5suspfixedpt} recovers the sphere free energy \eqref{SosS5susp}.
In terms of this quantity and of the prepotential \eqref{prepotential}, the fixed point contribution to the action can be rewritten as
\begin{equation}\label{eq:IS6Fversion}
    I_\pt=\frac{9}{4}F_{S^3}\frac{\mathcal{F}(b_i^\pt)}{\varepsilon_1^\pt\varepsilon_2^\pt}\,.
\end{equation}
This is the gravitational block term in \cite{Hosseini:2022vho}.
%%%%%%%%%%~~~~~~~~~~~~~~~~~~~~~~~~~~~~~~~~~~~~~~~~~~~~~~~~~~~~~~~~~~~~~~~~~~
%%%%%                       Spindle
%%%%%%%%%%~~~~~~~~~~~~~~~~~~~~~~~~~~~~~~~~~~~~~~~~~~~~~~~~~~~~~~~~~~~~~~~~~~
\paragraph{Spindle}
In particular we can consider an accelerating black hole, whose horizon is a spindle $\mathbb{WCP}^1_{[n_+,n_-]}$. The spacetime has topology AdS$_2\times\Sigma$ and there are two pairs of fixed points: the 2 poles of the suspensions ($N,S$) at the 2 poles of the spindle ($\pm$). Using that the on-shell action in \eqref{eq:IS6finalexp} is homogeneous of degree zero under rescaling all of the weights we may choose to set the weight on AdS$_2$ to be $\varepsilon_1=-1$. 
After the rescaling the weights on the spindle poles are $\varepsilon_2^{\pm}\equiv \mp\varepsilon/n_{\pm}$ and the charge condition \eqref{eq:charge4+6} then reads
\begin{align}
   -\frac{\varepsilon}{n_+}+\sum_{i=1}^3b_i^+=1\,,  \qquad
   \sigma\frac{\varepsilon}{n_-}+\sum_{i=1}^3b_i^-=1\,,
\end{align}
where $\sigma=\pm1$ for the twist/anti-twist.
We can directly use \eqref{eq:IS6Fversion} at the two poles of the spindle (and introducing the appropriate orbifold factor) to obtain 
\begin{equation}
    I=\frac{9}{4}F_{S^3}\frac{\mathcal{F}(b_i^+)-\sigma\mathcal{F}(b_i^-)}{\varepsilon}\,.
\end{equation}
Note that upon extremization this generalizes the result of the universal spindle black hole in \cite{Couzens:2022yiv} which uplifts the spindle black hole solution of four-dimensional minimal gauged supergravity on the suspension. This is an example of a solution which would not currently be possible to study without our ten-dimensional method.

%%%%%%%%%%~~~~~~~~~~~~~~~~~~~~~~~~~~~~~~~~~~~~~~~~~~~~~~~~~~~~~~~~~~~~~~~~~~
%%%%%                      Fixed surface
%%%%%%%%%%~~~~~~~~~~~~~~~~~~~~~~~~~~~~~~~~~~~~~~~~~~~~~~~~~~~~~~~~~~~~~~~~~~

\subsection{Fixed surfaces}\label{sec:susfixedsurface}

Now we consider the contribution coming from a fixed surface $\Sigma$ of $\xi$.
The fixed point set then consists of two bolts: the copies of the fixed surface at the two poles of the suspension. 
The bolt has a normal line bundle $L=\mathcal{O}(m)$ in $M_4$, while
the fibration of the internal space over the fixed surface is characterized by the integers $p_i$, $i=1,2,3$ defined as: 
\begin{equation}\label{eq:c1SusS5}
    \int_{\Sigma^N} c_1(L)=\int_{\Sigma^S} c_1(L)\equiv m\,\quad \int_{\Sigma^N} c_1(\mathcal{L}_i)=-\int_{\Sigma^S} c_1(\mathcal{L}_i)\equiv p_i\,.
\end{equation}
Recall that to each fixed point we have a set of signs $\sigma$ depending on the eigenvalues of the spin generators of the spinors at the fixed point. Let $\sigma_4$ be the eigenvalue for the copy of $\mathbb{C}$ inside $M_4$ at the bolt. This is fixed for both poles. On the other hand the signs for the suspension should flip sign going from the north pole to the south. We may fix the signs at the north pole to all be $+1$ without loss of generality. Then the condition for the spinor to be well-defined imposes the condition:
\begin{equation}\label{eq:c14+6bolt}
    \sigma_4 \Big[m +\chi(\Sigma)\Big]=p_1+p_2+p_3\, .
\end{equation}
In addition we have that the weights satisfy
\begin{equation}\label{eq:weights4+6bolt}
    \sigma_4 \varepsilon+\sum_{i=1}^{3}b_i=0\, .
\end{equation}

Having fixed the geometry let us quantize the fluxes. The cycles discussed in the previous section are still present, but there are now additional ones involving the fixed surface. As in the previous section the two-form and four-form fluxes through the compact cycles in the $S^6$ are again trivial
\begin{equation}\label{eq:SUSSigmahom}
        \int_{S_i^2}F^2=0 \,,\quad
        \int_{S_i^4}F^4=0
    \quad \implies \quad  \Phi_0^N=-\Phi_0^S=-\frac{1}{2}F_0(\y_N-\y_S)\,,
\end{equation}
while the $F_6$ flux through the $S^6$ remains unchanged
\begin{align}
    N&=\frac{1}{(2\pi\ls)^5}\int_{X_6}F_6=\frac{1}{b_1b_2b_3}\frac{n_0}{96\ls^6\pi^3}(\y_N-\y_S)^3\,.
\end{align}
Once again we obtain an expression for $\y_N$ as one of the cubic roots of this flux, giving \eqref{YNS5susp}.

We now turn to the ``new" cycles involving the bolt. 
We  have the two-cycle which is the bolt at the north and south pole of $S^6$ denoted by $[\Sigma^N]$ and $[\Sigma^S]$ respectively. We also have four-cycles which are given by the linearly embedded spheres inside $S^6$ fibred over the Riemann surface. These cycles are trivial in homology and the flux through them should vanish. 
Finally we have new six-cycles given by the linearly embedded four-spheres fibred over the Riemann surface. For the $S^4_i$ defined by $z_i=0$ inside $S^6$, the six-cycle satisfies $[S^4_i\ltimes \Sigma]=p_i [S^6]$ and we have a total of three such six-cycles.

The homology relations of the four-cycles $[S^2_i\ltimes\Sigma]$
impose 
\begin{equation}
    0=\int_{S^2_i\ltimes \Sigma}F_4=\frac{2\pi}{b_i}\left[\int_{\Sigma^{N}}\left(\Psi^{F_4}_2-\frac{2\pi}{b_i}\Psi^{F_4}_0 c_1(\mathcal{L}_i)\right)-\int_{\Sigma^{S}}\left(\Psi^{F_4}_2+\frac{2\pi}{b_i}\Psi^{F_4}_0 c_1(\mathcal{L}_i)\right)\right]\,,
\end{equation}
which after some simplification and using \eqref{eq:SUSSigmahom} gives the simple condition
\begin{equation}
    \int_{\Sigma^N}(\y_N F_2+\Phi_2)=\int_{\Sigma^S}(\y_S F_2+\Phi_2)\, .
\end{equation}
Since the two copies of $\Sigma$ are homologous, $[\Sigma^N]=[\Sigma^S]$, and $F_2$ is closed we have
\begin{equation}
   \frac{1}{2\pi\ls} \int_{\Sigma^N}F_2=   \frac{1}{2\pi\ls} \int_{\Sigma^S}F_2\equiv M\, .
\end{equation}
Then we may use this to eliminate the integral of $\Phi_2$ over $\Sigma^N$ as\footnote{Since $\Phi_2$ is not closed generically these need not be equal and the difference is parametrized by $M$.}
\begin{equation}
\int_{\Sigma^N}\Phi_2=\int_{\Sigma^S}\Phi_2+(\y_S-\y_N)M\, .
\end{equation} 
Finally the six-cycle homology relations impose
\begin{align}
    p_iN&=\frac{1}{(2\pi\ls)^5}\int_{S^4_i\ltimes \Sigma}F_6 \\
    &=\frac{1}{(2\pi\ls)^5}\frac{2\pi}{b_jb_k}\left[\int_{\Sigma^{N}}\Psi^{F_6}_2-\int_{\Sigma^{S}}\Psi^{F_6}_2-\left(2\pi\Psi^{F_6}_{0,N}-2\pi\Psi^{F_6}_{0,S}\right)\left( \frac{p_j}{b_j}+ \frac{p_k}{b_k}\right)\right]\,,
\end{align}
for $i=1,2,3$ with $i,j,k$ all distinct in the above formula. We may solve these as
\begin{align}
    \int_{\Sigma^S}\Phi_2&=\frac{\pi(\y_N-\y_S)}{6}\bigg[6\ls M +F_0(\y_N-\y_S) \sum_{i=1}^{3}\frac{p_i}{b_i}\bigg]\, .
\end{align}

This exhausts the homology relations and we can now compute the action:
\begin{align}
    I_\Sigma&=\frac{-\ii}{2(2\pi)^7\ls^8}\frac{(2\pi)^4}{\varepsilon b_1b_2b_3}\left[\int_{\Sigma^{N}}\Psi_2-\int_{\Sigma^{S}}\Psi_2-\left(2\pi\Psi_{0,N}-2\pi\Psi_{0,S}\right)\left( \frac{m}{\varepsilon}+\sum_{i=1}^3\frac{p_i}{b_i}\right)\right]\,.
\end{align}
In addition to the weights, the action still depends on $\y_S$ which should be extremized over. Performing the extremization for $\y_S$ we have two possible options, either $m=0$ (which forces $M=0$ and the dependence in $\y_S$ drops out) or for $m\neq 0$ $\y_S$ satisfies
\begin{equation}
    \y_S=- \y_N-\frac{4\pi \ls^2 M \varepsilon}{m n_0}\, .
\end{equation}
For this second, more general solution, we find
\begin{align}
    I_{\Sigma}=\frac{-\ii \me^{10 \pi \ii n_N/3} 3^{5/3}\pi}{2^{2/3}\times 5}\frac{1}{\varepsilon} \Big[(b_1 b_2 b_3)^{2/3} \frac{m}{\varepsilon}- \sum_{i=1}^{3} p_i \partial_{b_i}(b_1 b_2 b_3)^{2/3}\Big]  n_0^{1/3}N^{5/3}+  \frac{\ii \pi M^2 N}{m n_0}\, .
\end{align}
Note that setting $m=M=0$ this recovers the $m=0$ branch above which just drops the final term.
In terms of the $S^3$ free energy (for $m\neq 0$) we find
\begin{equation}
    I_{\Sigma}=\frac{9}{4}F_{S^3}\frac{1}{\varepsilon}\bigg[\sum_{i=1}^{3}p_i \partial_{b_i}\mathcal{F}(b_i)-\mathcal{F}(b_i) \frac{m}{\varepsilon}\bigg]+\frac{\ii\pi  M^2 N}{m n_0}\, ,
\end{equation}
where $\mathcal{F}(b_i)=(b_1 b_2 b_3)^{2/3}$ as in equation \eqref{prepotential}. Note that the term involving $M$ is purely imaginary and assuming that $M$ does not scale with $N$ is subleading in the large $N$ limit. {Notice that this has the same scaling as the subleading terms appearing in \cite{Hong:2021bsb}, (cf. equation (2.38) for example).  Similarly to the previous section, there can be additional subleading contributions arising from higher-derivative corrections and perturbative one-loop contributions.}

%%%%%%%%%%~~~~~~~~~~~~~~~~~~~~~~~~~~~~~~~~~~~~~~~~~~~~~~~~~~~~~~~~~~~~~~~~~~
%%%%%                       Black saddle
%%%%%%%%%%~~~~~~~~~~~~~~~~~~~~~~~~~~~~~~~~~~~~~~~~~~~~~~~~~~~~~~~~~~~~~~~~~~

\paragraph{Black bolt}

To exemplify our results consider taking $M_4$ to be a black saddle with Riemann surface horizon and let $m$ define the twisting of the thermal circle over the Riemann surface making it into a ``black bolt'', \cite{Benini:2017oxt, Azzurli:2017kxo, Hosseini:2017fjo}. From the perspective of the four-dimensional black hole the $p_i$ are the magnetic charges of three U$(1)$ gauge fields. We fix $\sigma_4=-1$ in the following so that \eqref{eq:c14+6bolt} and \eqref{eq:weights4+6bolt} read
\begin{equation}
    \chi(\Sigma_g)+m+\sum_{i=1}^{3}p_i =0 \, ,\qquad -\varepsilon+\sum_{i=1}^{3} b_i=0\, .
\end{equation}
It is useful to make the redefinition
\begin{equation}
    b_i=\varepsilon \Delta_i\, ,
\end{equation}
so that the action reads:\footnote{We have set the two-form flux term to vanish.}
\begin{equation}
I_{\text{black saddle}}=\frac{9}{4} F_{S^3}\Big[\sum_{i=1}^{3}p_i\partial_{\Delta_i}\mathcal{F}(\Delta)- m \mathcal{F}(\Delta)\Big]\, ,
\end{equation}
with $\mathcal{F}(\Delta)=(\Delta_1\Delta_2\Delta_3)^{2/3}$. 

%%%%%%%%%%~~~~~~~~~~~~~~~~~~~~~~~~~~~~~~~~~~~~~~~~~~~~~~~~~~~~~~~~~~~~~~~~~~
%%%%%                       Conclusion
%%%%%%%%%%~~~~~~~~~~~~~~~~~~~~~~~~~~~~~~~~~~~~~~~~~~~~~~~~~~~~~~~~~~~~~~~~~~

\section{Conclusion}

In this paper we have constructed equivariantly closed polyforms for the Page fluxes and the on-shell action of both type II supergravity theories directly in ten dimensions. Our construction relies only on the universal supersymmetry conditions of generalized geometry \cite{Tomasiello:2011eb} and therefore applies to arbitrary minimally supersymmetric type IIA and type IIB backgrounds admitting a pair of Killing spinors. This provides a unified localization framework in which both flux quantization and the on-shell action can be computed for arbitrary backgrounds. 

We illustrated the formalism for two large classes of massive type IIA solutions. The first was a four-dimensional hemisphere bundle over a six-dimensional external spacetime which generalizes the localization of Romans supergravity in \cite{Couzens:2025ghx} to the U$(1)^2$ theory. We derived the master formula \eqref{Ifinal} for the on-shell action for a general topology with arbitrary number of isolated fixed points, fixed surfaces and four-dimensional fixed submanifolds. 
Similarly, we analysed the suspension of a toric Sasaki--Einstein five-manifold fibred over a four-dimensional spacetime,  again obtaining a master formula for the on-shell action in \eqref{Ifinal4d}. The master formula depends on a prepotential which encodes the dependence on the fibres. In particular the prepotential is a power of the Sasakian volume of the Sasaki--Einstein manifold.  
In each case the localized action is expressed entirely in terms of the equivariant weights and the topology of the fixed point sets and makes no assumptions about a consistent truncation on the fibre.

One of the main advantages of the ten-dimensional perspective is that the localization computation completely bypasses the need for a consistent truncation. This represents a substantial simplification of the problem. Indeed, for a general Sasaki--Einstein suspension there is no known consistent truncation containing three vector fields and so obtaining these results otherwise seems prohibitively difficult. Furthermore, for some questions, one needs to work in ten dimensions and a gauged supergravity description is not sufficient. For example, to compute the on-shell action of probe branes one needs to work directly in ten dimensions to understand the embedding of the brane in the geometry \cite{BenettiGenolini:2026hmz}. Moreover, one could consider solutions where the external spacetime has punctures, i.e.\ a punctured Riemann surface. One can resolve these singularities, but this however requires the ten-dimensional embedding and the gauged supergravity description is insufficient; see, for example, \cite{Couzens:2025nxw,Couzens:2026qne}.

There are several natural directions for future work. First, while we have focused on massive type IIA, the construction of the equivariant polyforms is equally valid in type IIB, where it would be interesting to derive analogous master localization formulae. Second, in the massless limit our expressions should admit a direct interpretation in eleven-dimensional supergravity, providing a route towards localization in M-theory. Third, though we have considered geometries which have an asymptotically AdS factor the localization works equally well for other geometries. One could, for example, consider the on-shell action of non-conformal branes or asymptotically flat black holes. In these cases one needs to be more careful with boundary terms, which for the asymptotically AdS spacetimes we have considered (conjecturally) cancel. {Fourth, we only exploited the generalized geometrical structure to build the polyforms. However, this perspective should also help to gain further insight into the structure underlying our results, for example using the generalized equivariance of \cite{hu2007}.\footnote{{We thank the anonymous referee for bringing this to our attention.}}}
Finally, our analysis suggests that subleading contributions, arising from fluxes through resolution cycles, can also be incorporated systematically -- see \cite{Choi:2019dfu} for an example where such fluxes are incorporated in field theory. Understanding the physical interpretation of these corrections, as well as extending the formalism to more general geometries and flux backgrounds, would further clarify the scope of ten-dimensional localization.

\section*{Acknowledgments}

We thank Florian Gaar, Pietro Benetti Genolini, Chris Hull, Carlos N\'u\~nez, Tomás Ortín, Dmitri Sorokin and Peter West for discussions.
We thank the Centro de Ciencias de Benasque for hospitality in the final stages of this work.
JFS is supported in part by STFC grant ST/X000761/1.

%%%%%%%%%%%%%%%%%%%%%%%%%%%%%%%%%%%%%%%%%%%
%%%           Appendix
%%%%%%%%%%%%%%%%%%%%%%%%%%%%%%%%%%%%%%%%%%%
\appendix

%%%%%%%%%%%%%%%%%%%%%%%%%%%%%%%%%%%%%%%%%%%
%%%           
%%%%%%%%%%%%%%%%%%%%%%%%%%%%%%%%%%%%%%%%%%%

\section{Details of derivations}
In this appendix we present a more explicit derivation of the polyforms for the Page fluxes used in the main text, before giving the explicit polyforms in type IIB.

\subsection{Obtaining the polyforms}\label{app:polyforms}

 First consider the Maxwell equations in the democratic formalism \eqref{eq:FEOM} which we may rewrite in the form
\begin{equation}
    \dd_H \Fm=0\quad \Leftrightarrow \quad \dd(\me^{-B} \wedge \Fm)=0 \quad \Leftrightarrow \quad \dd F=0\, ,
\end{equation}
where the Page flux $F$ is defined in \eqref{eq:Pagedef}. Considering the contraction of $K$ into the Page flux, one finds
\begin{align}
    K\hook F&=-(K\hook B+\widetilde{K})\wedge F- \dd (\me^{-\phi}\mskip2mu\me^{-B}\mskip2mu\widehat{\Phi})\, .
\end{align}
Using the definition of $\y$ in equation \eqref{eq:ydef} and the definition of $\Phi$ in \eqref{eq:Phidef} we find
\begin{equation}
K\hook F=\dd [-(\y F+\Phi)]\equiv \dd \Psi_1\, .  
\end{equation}
This is the first step in constructing equivariantly closed forms with respect to $K$. Considering the contraction of $K$ into $\Phi$, we have
\begin{align}
    K\hook \Phi&=-(K\hook B)\wedge \Phi-\tilde{K}\wedge \Phi=-\dd \y \wedge \Phi\, ,
\end{align}
and therefore 
\begin{align}
K\hook \Psi_1&= \y \dd(\y F+\Phi)+\dd \y\wedge \Phi=\frac{1}{2}\dd (\y^2 F+2 \y \mskip2mu\Phi)%\\&
\equiv \dd \Psi_2\, .%\nonumber
\end{align}
Continuing in this manner we find that the full tower of equivariantly closed polyforms for the Page fluxes is given by 
\begin{equation}
    \Psi^F=\sum_{m=0} \Psi_m\, ,
\end{equation}
where 
\begin{align}
    \Psi_m &=\frac{(-1)^{m}}{m!}\Big(\y^{m} F+ m\y^{m-1}\Phi\Big )\, .
\end{align}
We have therefore constructed a polyform of equivariantly closed polyforms for all of the Page fluxes in both type II theories. For massive type IIA these read
\begin{align}\label{eq:IIAfluxpolyapp}
        \Psi^{F_0}&= F_0\, ,\\ \nn
        \Psi^{F_2}&= F_2-(\y F_0+\Phi_0)\, ,\\ \nn
        \Psi^{F_4}&=F_4-(\y F_2+\Phi_2)+\frac{1}{2}(\y^2 F_0+2 \y \Phi_0)\, ,\\ \nn
        \Psi^{F_6}&=F_6-(\y F_4+\Phi_4)+\frac{1}{2}(\y^2 F_2+2\y \Phi_2)-\frac{1}{3!}(\y^3 F_0+3\y^2 \Phi_0)\, ,\\ \nn
        \Psi^{F_8}&=F_8-(\y F_6+\Phi_6)+\frac{1}{2}(\y^2 F_4+2\y \Phi_4)-\frac{1}{3!}(\y^3 F_2+3\y^2 \Phi_2)
        +\frac{1}{4!}(\y^4 F_0+4 \y^3 \Phi_0)\, ,\\ \nn
        \Psi^{F_{10}}&=F_{10}-(\y F_8+\Phi_8)+\frac{1}{2}(\y^2 F_6+2\y \Phi_6)-\frac{1}{3!}(\y^3 F_4+3\y^2 \Phi_4)+\frac{1}{4!}(\y^4 F_2+4 \y^3\Phi_2) \\ \nn
         &\hspace{10.25cm}-\frac{1}{5!}(\y^5 F_0+5 \y^4 \Phi_0)\, .
\end{align}
For type IIB one finds that the equivariantly closed polyforms for each of the fluxes are
\begin{align}
    \Psi^{F_1}&=F_1\, ,\\
    \Psi^{F_3}&=F_3-(\y F_1+\Phi_1)\, ,\nonumber\\
    \Psi^{F_5}&=F_5-(\y F_3+\Phi_3)+\frac{1}{2}(\y^2 F_1+2 \y \Phi_1)\, ,\nonumber\\
    \Psi^{F_7}&= F_7-(\y F_5+\Phi_5)+\frac{1}{2}(\y^2 F_3+2 \y \Phi_3)-\frac{1}{3!}(\y^3 F_1+3 \y^2 \Phi_1)\, ,\nonumber\\
    \Psi^{F_9}&= F_9-(\y F_7+\Phi_7)+\frac{1}{2}(\y^2 F_5+2 \y \Phi_5)-\frac{1}{3!}(\y^3 F_3+3 \y^2 \Phi_3)+\frac{1}{4!}(\y^4 F_1+4 \y^3 \Phi_1)\, .\nonumber
\end{align}

\subsection{Putting the action partially on-shell}\label{app:action}

Starting from the pseudo action \eqref{pseudoaction}, the equation of motion for the dilaton is
\begin{equation}
     R-\frac{1}{2}|H|^2=4|\dd \phi|^2 -4\nabla^2\phi\, ,
\end{equation}
and we may go partially on-shell by eliminating the Ricci scalar. The partially on-shell action becomes 
\begin{align}
    S&=-\frac{1}{4(2\pi)^7\ls^8}\int_{M_{10}} \sum_{k}\Fm_k\wedge\star \Fm_k+ \frac{1}{(2\pi)^7\ls^8}\int_{M_{10}}\dd^{10}x \sqrt{-g}\mskip2mu\nabla^2(\me^{-2\phi})\nonumber \\
    &=-\frac{1}{4(2\pi)^7\ls^8}\int_{M_{10}}\mp \langle \Fm, \Fm\rangle+\frac{1}{(2\pi)^7\ls^8}\int_{\partial M_{10}}\dd^9 x \sqrt{-h}\mskip2mu n^{\mu}\nabla_{\mu}(\me^{-2\phi})\,,
\end{align}
where $n^{\mu}$ is the normal to the boundary, $h$ the induced metric on the boundary and $\langle \cdot\,,\cdot \rangle$ denotes the Mukai pairing. To write the Mukai pairing we have used the self-duality condition \eqref{eq:selfdual} with $\star\mskip2mu\lambda(F)=\mp \lambda(\star \mskip2mu F)$ in IIA/IIB respectively. This should then be interpreted as a formal rewriting; we will come back to this momentarily. 
Observe that the full NS-NS sector is a total derivative term for the partially on-shell action and in the following we will neglect such boundary terms.\footnote{These boundary terms may be important in other setups. Their analysis is beyond the scope of this paper, but we intend to come back to these contributions in future work.} 

We still have a little work to relate this to our polyforms in the previous section. Fortunately for any two-form $\omega$ the Mukai pairing satisfies
\begin{equation}
    \langle \me^{\omega} G, \me^{\omega} \widetilde{G}\rangle= \langle  G, \widetilde{G}\rangle\, ,
\end{equation}
and therefore we may replace the Maxwell fluxes $\Fm$ with the Page fluxes $F$ in the above. Our final result, ignoring boundary contributions, is then
\begin{equation}\label{actionapp}
    S=-\frac{1}{4(2\pi)^7\ls^8}\int_{M_{10}}\mp\langle F,F\rangle\,,
\end{equation}
where the $-$ sign is for type IIA and the $+$ sign for type IIB. In this form the partially on-shell action can be equivariantly completed using our previous results, provided we interpret \eqref{actionapp} suitably using a choice of electric/magnetic polarization so as to obtain a non-zero result. This is discussed in the main text.

%%%%%%%%%%%%%%%%%%%%%%%%%%%%%%%%%%%%%%%%%%%
%%%           
%%%%%%%%%%%%%%%%%%%%%%%%%%%%%%%%%%%%%%%%%%%

\section{Spinor projection condition}\label{app:spinors}

In this appendix we want to analyse the constraints imposed on the weights and various Chern classes by the existence of spinors on the background.\footnote{The analysis here closely follows the analysis in \cite{Couzens:2025ghx}.} Since our examples in this paper are in type IIA we will focus on this only here and leave the type IIB analysis to future work. 
In this section in particular we need to be careful with the Wick rotation that we have imposed. Though Majorana--Weyl spinors exist in ten-dimensional Lorentzian spacetime they do not exist in the Euclidean counterpart. It is then natural to give up the Majorana condition and keep two complex spinors $\epsilon_i$. One now needs to make a choice in the bilinears we have constructed. In the original Lorentzian theory we have
\begin{equation}
    \widehat{\Phi}=32\mskip2mu \epsilon_1\otimes \overline{\epsilon_2}=32\mskip2mu \epsilon_1\otimes (\epsilon_2^T C)\, ,
\end{equation}
with $C$ the Clifford algebra intertwiner satisfying $\Gamma_A^{T}=-C\mskip2mu \Gamma_A C^{-1}$. In the Euclidean theory it is then natural to work with $\widehat{\Phi}=32 \mskip2mu\epsilon_1\otimes (\epsilon_2^T C)$ working on a ``holomorphic slice" of the theory, see for example \cite{Bergshoeff:2007cg,DHoker:2025nid}.

Let us introduce a frame which is regular on the fixed point set and assume that the normal bundle decomposes as the direct sum of copies of $\mathbb{C}$. We may introduce rectangular coordinates $(x_I, y_I)$ on each copy  such that the Killing vector field takes the form
\begin{equation}
   K= \sum_{I=1}^{k} b_I (x_I \partial_{y_I}-y_I \partial_{x_I})\, .
\end{equation}
In this frame for a $10-2k$-dimensional fixed point set one can write 
\begin{equation}
    (\dd K)_{AB}= 2\left[\begin{pmatrix}
        0  & b_1\\
        -b_1 & 0
    \end{pmatrix}
    \oplus ...\oplus
    \begin{pmatrix}
        0  & b_k\\
        -b_k & 0
    \end{pmatrix} \oplus... \begin{pmatrix}
        0  & 0\\
        0 & 0
    \end{pmatrix}\right]_{AB}\,,
\end{equation}
where the $b_I$ are the weights of the action. Recall that the two spinors have vanishing Lie derivative along $K$ and this should hold in the Euclidean theory too, thus we have $\mathcal{L}_K\epsilon_i=0$. At a fixed point we have 
\begin{align}
    0& =\mathcal{L}_K\epsilon_{\alpha}|_{\text{f.p}}=K^{M}\nabla_{M} \epsilon_\alpha+\frac{1}{8}(\dd K)_{M N}\Gamma^{MN} \epsilon_\alpha|_{\text{f.p}}=\frac{1}{8}(\dd K)_{M N}\Gamma^{MN} \epsilon_\alpha|_{\text{f.p}}\nonumber\\ 
    & =-\frac{\ii}{2}\sum_{I=1}^{k}b_{I}J_{I}\epsilon_{\alpha}|_{\text{f.p}}\, ,
\end{align}
where we introduced the commuting spinor generators adapted to the frame:
\begin{equation}
    J_I=\ii \Gamma^{2I-1}\Gamma^{2I}\, ,\qquad J_I^2=1\, .
\end{equation}
We can decompose the spinors at the fixed points which gives us a set of signs 
\begin{equation}\label{projspinor}
    J_I \epsilon_{\alpha}\equiv \sigma_{I}^{(\alpha)}\epsilon_{\alpha}\, ,
\end{equation}
with $\sigma_{I}^{(\alpha)}=\pm 1$ and the condition becomes
\begin{equation}\label{eq:}
    \sum_{I=1}^{k} b_I \sigma_I^{(\alpha)}\epsilon_{\alpha}\Big|_{\text{f.p}}=0\, .
\end{equation}
There are {\it a priori} two sets of signs for each of the two spinors, but they in fact turn out to be related. To see this recall that in type IIA the polyform $\Phi$ has only even-dimensional forms, whilst in type IIB $\Phi$ contains only odd--dimensional forms. We have that the spin generators satisfy
\begin{equation}
    J_I^{T}C=-C J_I\, ,
\end{equation}
and therefore for the zero-form part of $\Phi$ in type IIA we have
\begin{equation}
    \Phi_0=\epsilon_2^T C \mskip2mu\epsilon_1=\epsilon_2^T C J_I^2 \epsilon_1=-\sigma_I^{(1)}\sigma_I^{(2)}\Phi_0\, .
\end{equation}
At a fixed point, obeying the projection condition \eqref{projspinor}, it follows that all polyforms in $\Phi$ are proportional to $\Phi_0$ and moreover that if $\Phi_0=0$ then either $\epsilon_1=0$ or $\epsilon_2=0$. The spinors should be nowhere vanishing and therefore at a fixed point we have that $\Phi_0$ is non-vanishing.
Since $\Phi_0$ is non-zero at the fixed points we have that  $\sigma_I^{(1)}=-\sigma_{I}^{(2)}\equiv \sigma_I$ for every direction in the normal bundle and we may choose, without loss of generality, to work with the signs for the spinor $\epsilon_1$. Notice that we must have $\prod_{I=1}^{5}\sigma_I=1$ in order for the spinors to have the correct chirality. Then, for massive type IIA, the weights at a $(5-k)$-dimensional fixed point satisfy
\begin{equation}\label{eq:genweightconstraint}
    \sum_{I=1}^{k} \sigma_{I}b_I=0\, .
\end{equation}

For the type IIB case the discussion is a little more subtle since one is necessarily dealing with forms and the basis is not invariant under the rotation and so we postpone this for future work.

%%%%%~~~~~~~~~~~~~~~~~~~~~~~~~~~~~~~~~~~~~~~~~~~~~~~~~~~~~~~~~

\paragraph{Isolated fixed point}

For isolated fixed points we have therefore shown that the weights necessarily satisfy the constraint
\begin{equation}\label{eq:fpweightconstraint}
    \sum_{I=1}^{5} \sigma_I b_I=0\, .
\end{equation}

%%%%%~~~~~~~~~~~~~~~~~~~~~~~~~~~~~~~~~~~~~~~~~~~~~~~~~~~~~~~~~

\paragraph{Two-dimensional fixed point sets}

Let us now consider restricting the spinors to a connected two-dimensional component of the fixed point set, $\mathscr{F}_2$. If we choose an orientation on $\mathscr{F}_2$ this forces $\mathscr{F}_2$ to be a Riemann surface, classified by its genuss\footnote{One can also consider punctures but we will restrict to smooth spaces here.} $\mathscr{F}_2=\Sigma_g$. Complex line bundles over the Riemann surface are uniquely fixed up to isomorphism by the first Chern number of the bundle, known as the degree, and as such we can unambiguously use the notation $\mathcal{O}(n)$ for the line bundle of degree $n$ over $\mathscr{F}_2=\Sigma_g$. The tangent space splits as 
\begin{equation}
    TM_{10}|_{\mathscr{F}_2}=T\Sigma_g \oplus \mathcal{N}\Sigma_g=T\Sigma_g \oplus\mathcal{L}_1\oplus ...\oplus \mathcal{L}_4=\mathcal{O}(\chi(\Sigma_g))\oplus \mathcal{O}(-p_1)\oplus...\oplus \mathcal{O}(-p_4)\, ,
\end{equation}
with $-p_I=\int_{\Sigma_g}c_1(\mathcal{L}_I)$, $I=1,..,4$ and $\chi(\Sigma_g)=2(1-g)$ the Euler characteristic of the Riemann surface. The chiral spinors satisfying the projection condition \eqref{projspinor} are sections of the complex line bundles
\begin{equation}\label{fatbundle}
    \mskip10mu
    \mathcal{O}\bigg[\pm \tfrac{1}{2}\bigg(-\sum_{I=1}^{4}\sigma_{I} p_I+\sigma_{\Sigma_g} \chi(\Sigma_g)\bigg)\bigg]\, ,
\end{equation}
where $\sigma_{\Sigma_g}=\prod_{I=1}^{4}\sigma_I$.
These two spinors should both be nowhere zero\footnote{The spinors satisfy first order differential equations, where a standard argument shows that if they vanish at a point they will vanish identically. But then all forms and fluxes vanish.} which implies that they both define nowhere-zero sections of the complex line bundle \eqref{fatbundle}. This is only possible if the line bundle is trivial. Therefore we necessarily find that for a two-dimensional fixed point set we have:
\begin{equation}
    \sum_{I=1}^{4}\sigma_I p_I =\sigma_{\Sigma_g} \chi(\Sigma_g)\, .
\end{equation}

%%%%%~~~~~~~~~~~~~~~~~~~~~~~~~~~~~~~~~~~~~~~~~~~~~~~~~~~~~~~~~
\paragraph{Four-dimensional fixed point set}

Consider now a four-dimensional fixed point set and let us denote such a connected component as $B_4$. In this case the tangent space decomposes as
\begin{equation}
    TM_{10}|_{B_4}= TB_4 \oplus \mathcal{L}_1\oplus \mathcal{L}_2\oplus \mathcal{L}_3\, .
\end{equation}
The three weights satisfy 
\begin{equation}
    \sum_{I=1}^{3} b_I \sigma_I=0\, .
\end{equation}
Let $\mathcal{S}^{\pm}_{B_4}$ be the two spinor bundles on $B_4$; then the ten-dimensional spinors satisfying the projection condition \eqref{projspinor} are sections of the rank-two spinor bundle
\begin{equation}
    \mathcal{E}=\mathcal{S}^{\eta}_{B_4}\otimes \mathcal{L}_1^{\pm\sigma_1/2}\otimes \mathcal{L}_2^{\pm\sigma_2/2}\otimes \mathcal{L}_3^{\pm\sigma_3/2}\, ,
\end{equation}
where $\eta=\pm$ is the chirality of the spinor on $B_4$, and is not correlated with the signs in the exponents  but rather the product of the $\sigma$'s: $\prod_{I=1}^{3}\sigma_I=\eta$. As before, since the spinors are nowhere-vanishing sections of a bundle we have that 
\begin{equation}
    \int_{B_4}c_2(E)=0\, .
\end{equation}
The spinor spans a trivial line subbundle of $\mathcal{E}$ and therefore we have $\mathcal{E}\cong \mathcal{L}_{\epsilon}\oplus \mathcal{L}_{\perp}$. Let us now assume that $B_4$ admits a U$(2)$ structure, for example induced by supersymmetry. In terms of this almost complex structure the spinor bundles on $B_4$ are given by 
\begin{equation}
    \mathcal{S}^{+}_{B_4}=K_{B_4}^{1/2}\oplus K_{B_4}^{-1/2}\, ,\quad \mathcal{S}^{-}_{B_4}=K_{B_4}^{1/2}\otimes \Lambda^{(0,1)}(B_4)\, ,
\end{equation}
where $K_{B_4}$ is the canonical line bundle and $\Lambda^{(0,1)}$ denotes the set of $(0,1)$-forms on $B_4$. With our assumption of a U$(2)$ structure on $B_4$,  $\mathcal{S}^{+}_{B_4}$ decomposes into the sum of line bundles; however, in general $\mathcal{S}^{-}_{B_4}$ does not decompose this way unless $\Lambda^{(0,1)}$ does -- see \cite{Couzens:2025ghx} for some further discussion. Therefore in the case of a positive chirality spinor on $B_4$ we have
\begin{equation}
    \mathcal{E}=\Big(K_{B_4}^{1/2}\otimes \mathcal{L}_1^{\sigma_1/2}\otimes \mathcal{L}_2^{\sigma_2/2}\otimes \mathcal{L}_3^{\sigma_3/2}\Big)\oplus  \Big(K_{B_4}^{-1/2}\otimes \mathcal{L}_1^{\sigma_1/2}\otimes \mathcal{L}_2^{\sigma_2/2}\otimes \mathcal{L}_3^{\sigma_3/2}\Big)
\end{equation}
and one of these line bundles is necessarily trivial. Therefore we have that 
\begin{equation}\label{eq:4df.p:c1}
    \pm c_1(K_{B_4})=\sum_{I=1}^{3}\sigma_I c_1(\mathcal{L}_I)\, ,
\end{equation}
with the $\pm$ sign the choice of canonical or anti-canonical bundle. 

%%%%%~~~~~~~~~~~~~~~~~~~~~~~~~~~~~~~~~~~~~~~~~~~~~~~~~~~~~~~~~

\paragraph{Six-dimensional fixed point sets}

We now analyse a six-dimensional fixed point set, $B_6$.
The tangent space decomposes as 
\begin{equation}
    TM_{10}|_{B_6}=TB_6 \oplus \mathcal{L}_1\oplus \mathcal{L}_2\, .
\end{equation}
The Killing vector rotates the two line bundles and the weights necessarily satisfy 
\begin{equation}
    \sigma_1 b_1+\sigma_2 b_2=0\, .
\end{equation} The spinors satisfying \eqref{projspinor} are now sections of the rank 4 bundle
\begin{equation}
\mathcal{E}=S^{\pm \sigma_1\sigma_2}_{B_6}\otimes \mathcal{L}_1^{\pm \sigma_1/2}\otimes \mathcal{L}_2^{\pm \sigma_2/2}\, .
\end{equation}
Notice that necessarily one of the spinors involves a positive chiral spinor on $B_6$ and the other involves a spinor of the other chirality. Assuming that $B_6$ admits an almost complex structure, with $K$ the canonical bundle, the positive and negative spin bundles on $B_6$ decompose as 
\begin{equation}
    S_{B_6}^{+}=K^{1/2}\otimes\big( \Lambda^{(0,0)}\oplus \Lambda^{(0,2)}\big)\, ,\quad S_{B_6}^{-}=K^{1/2}\otimes\big( \Lambda^{(0,1)}\oplus \Lambda^{(0,3)}\big)\, .
\end{equation}
Observing that $\Lambda^{(0,3)}=\overline{K_{B_6}}$, i.e. the anti-canonical bundle, we have that the spinors preserving the $U(3)$ structure on $B_6$ are nowhere-vanishing sections of 
\begin{equation}
    K^{ \pm \sigma_1 \sigma_2/2}\otimes \mathcal{L}_1^{\pm \sigma_1/2}\otimes \mathcal{L}_2^{\pm \sigma_2/2}\, ,
\end{equation}
and therefore we find that we necessarily have
\begin{equation}
    \sigma_1 \sigma_2 ~c_1(K_{B_6})+\sum_{I=1}^{2} \sigma_I c_1(\mathcal{L}_I)=0\, .
\end{equation}
There may be more general choices involving spinors in $\Lambda^{(0,1)}$ and $\Lambda^{(0,2)}$, in analogy with the discussion in the four-dimensional fixed point set example.

%%%%%~~~~~~~~~~~~~~~~~~~~~~~~~~~~~~~~~~~~~~~~~~~~~~~~~~~~~~~~~
\paragraph{Eight-dimensional fixed point sets}

There is an immediate obstruction to having an eight-dimensional fixed point: the invariance of the Killing spinor under $K$ implies that the solitary weight must vanish. This then implies that the action is trivial in the normal direction and thus there is no fixed point in this case. A similar result holds for four-dimensional ungauged supergravity, for precisely the same reason.\footnote{Notice that in the 
minimal AdS$_2$ classification in \cite{Legramandi:2023fjr} the bilinear $K$ is tangent to the AdS$_2$ factor. 
In Poincar\'e coordinates $\dd s^2_{\text{AdS}_2}=-\rho^2\diff t^2 + \diff\rho^2/\rho^2$ this Killing vector bilinear is $K=\partial_t$, under which the corresponding Killing spinor is 
indeed uncharged. Note that $\partial_t$ has no fixed points (in the Wick rotation), so this is not in contradiction with the above comments.
}

%%%%%~~~~~~~~~~~~~~~~~~~~~~~~~~~~~~~~~~~~~~~~~~~~~~~~~~~~~~~~~

%%%%%%%%%%%%%%%%%%%%%%%%%%%%%%%%%%%%%%%%%%%
%%%          
%%%%%%%%%%%%%%%%%%%%%%%%%%%%%%%%%%%%%%%%%%%

\section{Homology considerations}\label{app:homology}

In this appendix we prove the homology relations used in section \ref{sec:6+4}. The extension of these arguments to the relations in section \ref{sec:4+6} is straightforward.

The manifold $M_{10}$ has a ``boundary'' which is  $\partial HS^4\rightarrow M_6$. 
We denote this by $\partial\subset M_{10}$, noting that $M_6$ itself will 
have a (conformal) boundary which is different.\footnote{Thus $\partial$ itself has a boundary, namely $S^3\rightarrow \partial M_6$.} Then in what follows we will 
take the RR Page forms 
\begin{align}\label{Fbcs}
F_{2n}\in H^{2n}(M_{10},\partial)\, ,
\end{align}
 in particular 
meaning they have zero restriction to the D8/O8 boundary $\partial$, so 
$F_{2n} |_\partial = 0$. 
One can regard this as a choice of boundary conditions, where we note this 
is satisfied for uplifts of the minimal Romans theory (the RR forms in (2.1) of \cite{Couzens:2025ghx} 
are all uplifts of Romans fields that are multiplied by functions 
that go to zero on $\partial HS^4$). We will need also the stronger condition 
that also the lower degree parts of equivariantly closed forms for these fluxes are zero -- concretely, that $\y$ and the bilinear form $\Phi$ are zero when restricted to 
$\partial$. To see this one should solve the 
ten-dimensional killing spinor equations near a supersymmetric D8/O8 boundary. Using the results in \cite{Passias:2012vp} we can see that indeed the spinors vanish on the D8/O8 locus and therefore both $\y$ and $\Phi$ do too. We will assume that
\begin{align}\label{Psibcs}
\Psi^{F_{2n}}\in H_K(M_{10},\partial)\, ,
\end{align}
with \eqref{Fbcs} then a corollary of this, where $H_K$ denotes equivariant cohomology for $\diff_K$.

As a first consequence,
it is immediate from \eqref{Psibcs} that there is no flux of the Page RR two-form $F_2$
through the copies of $HS^2_i$ at the fixed points. Indeed,
  since $H^2(HS^4,S^3)=0$, necessarily $F_2=\diff\omega$ is 
exact restricted to a fibre $HS^4$ where $\omega |_{S^3}=0$, but then 
$\int_{HS^2_i}F_2 = \int_{S^1_i}\omega = 0$ where $S^1_i\subset S^3$ lies on the boundary. 

Next,
consider the 6-manifold $B_6$ which is $HS^4$ fibred over a two-dimensional manifold 
$\Sigma$, with four-cycles $C_i$ which are $HS^2_i$ fibred over $\Sigma$. 
These and the fibre $HS^4$ are classes in $H_4(B_6,\partial,\Z)$. 
Consider a closed four-form $F_4\in H^4(B_6,\partial)$. 
Fix, say, $C_1$, and let the Poincar\'e dual be $\Psi_1$, which has support
along the fibre of $\mathcal{L}_2$. This satisfies $\Psi_1=\pi^*c_1(\mathcal{L}_2)\in H^2(B_6)$ by explicit construction of this Poincar\'e dual, where $\pi:HS^4\rightarrow\Sigma$ is the projection. 
 Then
\begin{align}
N_1 \equiv \int_{C_1}F_4  = \int_{B_6}F_4\wedge \Psi_1 =  \int_{B_6}
F_4 \wedge \pi^*c_1(\mathcal{L}_2) = N \int_\Sigma c_1(\mathcal{L}_2)\, ,
\end{align}
where $N\equiv \int_{HS^4}F_4$ is the integral of $F_4$ over the 
$HS^4$ fibre. 

Finally, consider $HS^4$ fibred over $B_4$, and let $B_4^*$ be the copy 
of $B_4$ at the pole of $HS^4$. The Poincar\'e dual of $B_4^*$ inside the total space $B_8$ is 
$\Psi_1\wedge \Psi_2=\pi^*[c_1(\mathcal{L}_1)\wedge c_1(\mathcal{L}_2)]\in H^4(B_8)$. Then
\begin{align}
\int_{B_4^*}F_4 = \int_{B_8}F_4 \wedge \Psi_1\wedge \Psi_2 
= \int_{B_4}  c_1(\mathcal{L}_1)\wedge c_1(\mathcal{L}_2)\int_{HS^4_{\text{fibre}}} F_4  = \langle p_1, p_2\rangle N\, ,
\end{align}
with notation as in the main text.

%%%%%%%%%%%%%%%%%%%%%%%%%%%%%%%%%%%%%%%%%%%
%%%           
%%%%%%%%%%%%%%%%%%%%%%%%%%%%%%%%%%%%%%%%%%%

\section{Suspension over $T^{1,1}$}\label{app:T11}

In this section we will consider the suspension of $T^{1,1}$. Let $\{\partial_{\varphi_i}\}$ be a basis for the U$(1)^3$ action on $T^{1,1}$ adapted so that the Killing spinor is charged only under $\partial_{\varphi_1}$ and uncharged under the remaining two. In this basis $T^{1,1}$ is described by the four toric vectors
\begin{equation}
    v_1=\begin{pmatrix} 1\\0\\0
    \end{pmatrix}\, ,\quad v_2=\begin{pmatrix} 1\\0\\1
    \end{pmatrix}\, ,\quad v_3=\begin{pmatrix} 1\\1\\1
    \end{pmatrix}\, ,\quad v_4=\begin{pmatrix} 1\\1\\0
    \end{pmatrix}\, .
\end{equation}
These define the facets of a three-dimensional cone. 
The tip of the cone is not smooth, where one can see this by noting that there are four faces meeting at the tip rather than three. One manifestation of the singularity is that the normal bundle of the fixed point does not decompose into the sum of three line bundles, and this makes the localization  more subtle. A simple way to proceed is to perform a small resolution of the singular point, blowing up a $\mathbb{CP}^1$ at the tip, and then later collapsing the blow-up. Since this is a crepant resolution this does not change the volume computed using localization \cite{Martelli:2006yb}, and thus the result does not depend on the resolution at the end. 

\begin{figure}

\begin{subfigure}{0.32\textwidth}
    \begin{center}
        \begin{tikzpicture}

%For the faces
\coordinate (origin) at (0,0);
\coordinate (v1) at (1.2,-2.5);
\coordinate (v3) at (-1.5,-2.5);
\coordinate (v2) at (0.45,-3);
\coordinate (v4) at (-0.7,-2);

%Vectors
\coordinate (vec1s) at (0.4,-1.3);
\coordinate (vec1e) at (1.1,-0.76);
\coordinate (vec1en) at (1.2,-0.66);

\coordinate (vec2s) at (0.1,-1.1);
\coordinate (vec2e) at (0.5,-0.36);
\coordinate (vec2en) at (0.65,-0.3);

\coordinate (vec3s) at (-0.58,-1.25);
\coordinate (vec3e) at (-1.1,-0.76);
\coordinate (vec3en) at (-1.23,-0.62);

\coordinate (vec4s) at (-0.3,-1.3);
\coordinate (vec4e) at (-0.4,-0.76);
\coordinate (vec4en) at (-0.16,-0.66);

\draw[->,red] (vec2s)--(vec2e);
\draw[->,red] (vec3s)--(vec3e);
%Face lines
\draw[-] (origin)--(v1);
\draw[-] (origin)--(v2);
\draw[-] (origin)--(v3);
\draw[dashed] (origin)--(v4);
\draw[dashed] (v3)--(v4)--(v1)--(v2)--(v3);

% Arrows
\draw[->,red] (vec1s)--(vec1e);
\draw[->,red] (vec4s)--(vec4e);

\node at(vec1en) {$v_1$};
\node at(vec2en) {$v_2$};
\node at(vec3en) {$v_3$};
\node at(vec4en) {$v_4$}; 
        \end{tikzpicture}
         \caption{The cone of $T^{1,1}$. \\ $\quad$ }
    \end{center}
    
    \end{subfigure}
\begin{subfigure}{0.32\textwidth}
    \begin{center}
        \begin{tikzpicture}

\coordinate (origin1) at (-0.5,0.2);
\coordinate (origin2) at (0.5,0);
\coordinate (v1) at (1.2,-2.5);
\coordinate (v3) at (-1.5,-2.5);
\coordinate (v2) at (0.45,-3);
\coordinate (v4) at (-0.7,-2);

%Vectors
\coordinate (vec1s) at (0.6,-1.3);
\coordinate (vec1e) at (1.1,-0.76);
\coordinate (vec1en) at (1.2,-0.66);

\coordinate (vec2s) at (0.2,-0.8);
\coordinate (vec2e) at (0.85,-0.1);
\coordinate (vec2en) at (1.05,-0.05);

\coordinate (vec3s) at (-0.78,-1.25);
\coordinate (vec3e) at (-1.1,-0.76);
\coordinate (vec3en) at (-1.23,-0.62);

\coordinate (vec4s) at (-0.3,-1.3);
\coordinate (vec4e) at (-0.4,-0.76);
\coordinate (vec4en) at (-0.16,-0.66);

\draw[->,red] (vec2s)--(vec2e);
\draw[->,red] (vec3s)--(vec3e);

\draw[-] (origin2)--(v1);
\draw[-] (origin2)--(v2);
\draw[-] (origin1)--(v3);
\draw[dashed] (origin1)--(v4);
\draw[-] (origin1)--(origin2);

\draw[dashed] (v3)--(v4)--(v1)--(v2)--(v3);

% Arrows
\draw[->,red] (vec1s)--(vec1e);
\draw[->,red] (vec4s)--(vec4e);

\node at(vec1en) {$v_1$};
\node at(vec2en) {$v_2$};
\node at(vec3en) {$v_3$};
\node at(vec4en) {$v_4$};
            
        \end{tikzpicture}
         \caption{The first choice of small \\ resolution. }
    \end{center}
    \end{subfigure}
    \begin{subfigure}{0.32\textwidth}
    \begin{center}
        \begin{tikzpicture}

\coordinate (origin2) at (0.3,0.3);
\coordinate (origin1) at (-0.25,-0.2);
\coordinate (v1) at (1.2,-2.5);
\coordinate (v3) at (-1.5,-2.5);
\coordinate (v2) at (0.45,-3);
\coordinate (v4) at (-0.7,-2);

%Vectors
\coordinate (vec1s) at (0.4,-1.3);
\coordinate (vec1e) at (1.1,-0.76);
\coordinate (vec1en) at (1.2,-0.66);

\coordinate (vec2s) at (0.2,-0.8);
\coordinate (vec2e) at (0.85,-0.1);
\coordinate (vec2en) at (1.05,-0.05);

\coordinate (vec3s) at (-0.5,-1.25);
\coordinate (vec3e) at (-1.1,-0.76);
\coordinate (vec3en) at (-1.23,-0.62);

\coordinate (vec4s) at (-0.3,-1.8);
\coordinate (vec4e) at (-0.4,-1);
\coordinate (vec4en) at (-0.39,-0.85);

\draw[->,red] (vec2s)--(vec2e);
\draw[->,red] (vec3s)--(vec3e);

\draw[-] (origin2)--(v1);
\draw[-] (origin1)--(v2);
\draw[-] (origin1)--(v3);
\draw[dashed] (origin2)--(v4);
\draw[-] (origin1)--(origin2);

\draw[dashed] (v3)--(v4)--(v1)--(v2)--(v3);

% Arrows
\draw[->,red] (vec1s)--(vec1e);
\draw[->,red] (vec4s)--(vec4e);

\node at(vec1en) {$v_1$};
\node at(vec2en) {$v_2$};
\node at(vec3en) {$v_3$};
\node at(vec4en) {$v_4$};
            
        \end{tikzpicture}
    \end{center}
     \caption{The second choice of small resolution. }
    \end{subfigure}
   \caption{The cone of $T^{1,1}$ along with the two choices of small resolution.}\label{fig:res1}
\end{figure}
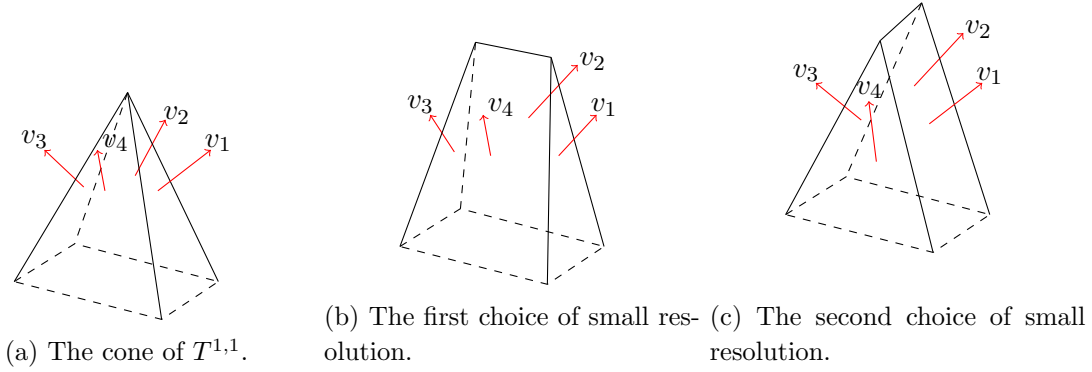

There are two options for the resolution (see figure~\ref{fig:res1}), related by a flop, specified by  taking either pair of maximal cones for the resolution
\begin{equation}\label{eq:smallres}
\begin{split}
    \tau_1^{+}&=\{v_1,v_2,v_3\}\, ,\quad \tau_1^{-}=\{v_1, v_3, v_4\}\, ,\\
    \tau_2^{+}&=\{v_1,v_2,v_4\}\, ,\quad \tau_2^{-}=\{v_2, v_3, v_4\}\, .
\end{split}
\end{equation}
To describe the suspension we need to glue two copies of the cone over $T^{1,1}$ together. We cut the (infinite) cones off at some radius and glue the two together, reversing the orientation of one copy and identifying the faces. For this to make sense we must ensure that the glued faces are described by the same toric vector, or more precisely the boundaries of the two cones are the same. The result need not be a (symplectic or complex) toric space in the usual sense. There are two distinct choices: one gives rise to $\mathbb{CP}^3$, whilst the other has the topology of a trivial $S^2$ bundle over $S^4$ and thus does not admit a Hamiltonian toric action. We will consider this choice and leave analysing $\mathbb{CP}^3$ for future work.

%%%%%%%%%%%%%%%%%%%%%%%%%%%%%%%%%%%%%%%%%%%
%%%           
%%%%%%%%%%%%%%%%%%%%%%%%%%%%%%%%%%%%%%%%%%%

\subsection{Same small resolution}

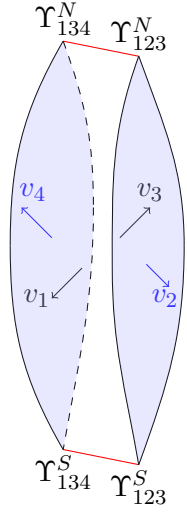
\begin{figure}
    \centering
  \begin{tikzpicture}

    %%Labelling the nodes. 
\coordinate (origin1) at (-0.5,0.2);
\coordinate (origin1lab) at (-0.5,0.5);

\coordinate (origin2) at (0.5,0);
\coordinate (origin2lab) at (0.5,0.3);

\coordinate (origin1S) at (-0.5,-5.2);
\coordinate (origin1Slab) at (-0.5,-5.5);

\coordinate (origin2S) at (0.5,-5.4);
\coordinate (origin2Slab) at (0.5,-5.7);

 %midpoints
\coordinate (v1) at (1.1,-2.5);
\coordinate (v3) at (-1.2,-2.5);
\coordinate (v2) at (0.15,-2.8);
\coordinate (v4) at (-0.1,-2);

% labelling the vertices
\node at (origin1lab) {$\y_{134}^{N}$};
\node at (origin2lab) {$\y_{123}^{N}$};
\node at (origin1Slab) {$\y_{134}^{S}$};
\node at (origin2Slab) {$\y_{123}^{S}$};

%drawing the bounding curves
\draw[-,red] (origin1)--(origin2);
\draw[-,red] (origin1S)--(origin2S);

\draw[out=240,in=90] (origin1) to (v3);
\draw[out=-90, in=120] (v3) to (origin1S);

\draw[dashed,out=290, in=90] (origin1) to (v4);
\draw[dashed,out=270, in=80] (v4) to (origin1S);

\draw[out=250, in=90] (origin2) to (v2);
\draw[out=270, in=100] (v2) to (origin2S);

\draw[out=290, in=90] (origin2) to (v1);
\draw[out=270, in=70] (v1) to (origin2S);

%Vector time

\coordinate (v2s) at (0.6,-2.75);
\coordinate (v2e) at (0.9,-3.05);
\coordinate (v2lab) at (0.85,-3.2);
\draw[->,blue] (v2s)--(v2e);
\node at (v2lab) {\textcolor{blue}{$v_2$}};

\coordinate (v1s) at (-0.25,-2.8);
\coordinate (v1e) at (-0.65,-3.2);
\coordinate (v1lab) at (-0.85,-3.2);
\draw[->] (v1s)--(v1e);
\node at (v1lab) {$v_1$};

\coordinate (v3s) at (0.25,-2.4);
\coordinate (v3e) at (0.65,-2.);
\coordinate (v3lab) at (0.65,-1.8);
\draw[->] (v3s)--(v3e);
\node at (v3lab) {$v_3$};

\coordinate (v4s) at (-0.65,-2.4);
\coordinate (v4e) at (-1.05,-2.);
\coordinate (v4lab) at (-0.9,-1.8);
\draw[->,blue] (v4s)--(v4e);
\node at (v4lab) {\textcolor{blue}{$v_4$}};

\fill[blue!30, opacity=0.3] (origin2) to [out=250, in=90] (v2) to [out=270, in=100] (origin2S) to [out=70,in=270] (v1) to [out=90, in=290] (origin2);

\fill[blue!30, opacity=0.3] (origin1) to [out=240, in=90] (v3) to [out=-90, in=120] (origin1S) to [out=80, in=270] (v4) to [out=90,in=290] (origin1);

    \end{tikzpicture}

    \caption{The ``toric" diagram for the same small resolution at both poles. The blue faces are $S^4$'s (not toric) whilst the two red lines are the resolution $\mathbb{CP}^1$'s which are homologous. The black lines denote $S^2_i\subset S^4$ at the poles of the resolution $\mathbb{CP}^1$, and the white faces are $\mathbb{CP}^1\times S^2_i$.}
    \label{fig:sameres}
\end{figure}

For our first example we will consider the same small resolution at both poles of the suspension. Without loss of generality we will take the first resolution in \eqref{eq:smallres}, with the second giving an equivalent result under relabelling. The final result is that $X_6=S^2\times S^4$ -- see figure~\ref{fig:sameres}. 

Let us consider the various cycles in $X_6$. 
We have the two blown-up spheres which are homologous; we label their homology class as $[S^2]$ in the following. We may also consider the two two-spheres, $[S^2_i]$ inside $S^4$ at a pole of the $S^2$. These two two-cycles are trivial in homology unlike the class $[S^2]$. There are two copies of the $S^4$ located at either pole of the $S^2$. These are homologous and we will denote this class as $[S^4]$. One can also construct the trivial four-cycles from the direct product of the $S^2$ and one of the trivial $[S^2_i]$ cycles. This gives all the cycles in the geometry. 

In order to collapse the blown-up cycle at the end of the computation we must require that the flux through the cycle $[S^2]$ vanishes. It would be interesting to keep this flux, which can be interpreted as turning on a magnetic baryonic flux; however, this makes finding the critical points a lot more challenging, and so for simplicity we set it to zero. 

To proceed we first impose the homology relations, starting with the two-cycles. It is useful to first define
\begin{equation}
    D_{123}\equiv(b_0-b_2)(b_2-b_1)b_1\, ,\qquad 
    D_{134}\equiv(b_0-b_1)(b_1-b_2)b_2\, ,
\end{equation}
which are the products of the weights at the fixed points. 
In terms of these weights the usual $T^{1,1}$ volume is given by 
\begin{equation}
    \vol(T^{1,1})=\frac{1}{D_{123}}+\frac{1}{D_{134}}=\frac{b_1}{b_2 b_3(b_1-b_2)(b_1-b_3)}\, .
\end{equation}
Let us also define
\begin{equation}
    Z\equiv-\Psi^{F_2}_0=\Phi_0+F_0 \y\, .
\end{equation}
Integrating the two-form flux over the trivial two-cycles imposes
\begin{equation}
    Z_{123}^{N}=Z_{123}^{S}\, ,\quad Z_{134}^{N}=Z_{134}^{S}\, ,
\end{equation}
whilst if we set the quantized flux through the resolution $\mathbb{CP}^1$ to be $B$ we have
\begin{equation}
    B\equiv \frac{1}{2\pi\ls}\int_{\mathbb{CP}^1}F_2\quad \Leftrightarrow \quad Z_{123}^N-Z_{134}^N= \ls (b_2-b_3) B=Z_{123}^S-Z_{134}^S\, .
\end{equation}
In the following we will set $B=0$ in order to blow down at the very end, which implies that in terms of the $Z$ variables the two-cycle homology relations imply that $Z$ is equal at all the fixed points.

To perform a similar analysis for the four-cycles we first define:
\begin{equation}\label{eq:ypmT11}
    \y_{123}^{\pm}\equiv\frac{1}{\pi\ls^2}(\y_{123}^{N}\pm \y_{123}^{S})\, ,\quad  \y_{134}^{\pm}\equiv\frac{1}{\pi\ls^2}(\y_{134}^{N}\pm \y_{134}^{S})\, ,
\end{equation}
and we define the one independent value of $Z$ to be $Z_0$, that is
\begin{equation}
     Z_0=Z_{123}^{N}=Z_{123}^{S}=Z_{134}^{N}=Z_{134}^{S}\, .
\end{equation}
Then the four-cycle homology relations through the cycles $[S^2\times S^2_i]$ gives\footnote{Keeping the two-form flux through the resolution $\mathbb{CP}^1$ non-trivial would change the relations here.}
\begin{equation}
    \y_{123}^+=\frac{4 Z_0}{n_0}=\y_{134}^{+}\, ,
\end{equation}
which also implies that the flux through the $S^4$ necessarily vanishes.\footnote{This is also true when we do not set $B=0$ and holds in general.} Finally, quantizing the $F_6$ flux through the internal space we find
\begin{equation}
    96 (b_2-b_3)N=n_0 \big[(\y_{134}^-)^3-(\y_{123}^-)^3\big]\, ,
\end{equation}
which we may solve for $\y_{123}^-$:
\begin{equation}
    \y_{123}^{-}=\me^{2\pi\ii n_{123}/3 }\bigg[\frac{96 (b_2-b_3)N+n_0 (\y_{134}^+)^3}{n_0}\bigg]^{1/3}\, ,
\end{equation}
where we have included the third root of unity with $n_{123}=\{0,1,2\}$. 

Up to this point, the analysis above applies to both an isolated fixed point or a bolt. We will first consider the former, dealing with an isolated fixed point before looking at the bolt computation. 

\paragraph{Isolated fixed point:}
The partially on-shell action now depends only on the free parameters $Z_0$, $\y_{134}^-$ and $\{b_1,b_2,b_3\}$. It is simple to see that the extremization over $Z_0$ imposes $Z_0=0$ whilst the extremization over $\y_{134}^-$ imposes
\begin{equation}
    \y_{134}^-= \me^{\pi \ii n_{134}/3}\y_{123}^-\, .
\end{equation}
That is, they are equal up to a sixth root of unity. In order for us to blow down we must take them to be the same. The final result is that the partially on-shell action for an isolated fixed point is
\begin{equation}
    I=\frac{9}{4}F_{S^3}^{T^{1,1}} \frac{\mathcal{F}}{\epsilon_1\epsilon_2}\, .
\end{equation}
To write the final result we have defined the prepotential
\begin{equation}
    \mathcal{F}(b_i)=\bigg(\frac{\vol_{T^{1,1}}}{\vol_{T^{1,1}}(\vec{b}\,)}\bigg)^{2/3}\, ,
\end{equation}
where $\vol_{T^{1,1}}(b_i)$ is the off-shell Sasakian volume and $\vol_{T^{1,1}}=\tfrac{16 \pi^3}{27}$ is the on-shell value.

\paragraph{Bolt contribution}
Consider now a bolt $\Sigma\subset M_4$. In $X_6$ we twist each of the three U$(1)$'s generated by $\partial_{\varphi_i}$ by a complex line bundle $\mathcal{L}_i$ on $\Sigma$ and also turn on a normal line bundle $L$ inside $M_4$ so that the total normal bundle  is $L\oplus_{i=1}^3\mathcal{L}_i\rightarrow\Sigma$.  Over the four copies of $\Sigma$ at the fixed points of the internal space we take
\begin{equation}
  \int_{\Sigma^N_{\pm}} c_1(L)=\int_{\Sigma^S_{\pm}} c_1(L)\equiv m\,,\qquad   \int_{\Sigma^N_{\pm}} c_1(\mathcal{L}_i)=-\int_{\Sigma^S_{\pm}} c_1(\mathcal{L}_i)=p_i\, ,
\end{equation}
as in \eqref{eq:c1SusS5}. The total space is then 
\begin{equation}
    M_{10}=\mathcal{O}(\{m,\vec{p}\})_{\Sigma_g}\times_{U(1)^4}\mathbb{C}\times X_6\,,
\end{equation}
i.e. we use the U$(1)^4$ transition functions of $\mathcal{O}(\{m,\vec{p}\})_{\Sigma_g}$ to fibre $\mathbb{C}\times X_6$ over $\Sigma_g$.

Since we have a bolt we now have additional two-, four- and six-cycles over which we can quantize the flux. We may turn on a two-form potential through the bolt and as in section~\ref{sec:susfixedsurface} we take
\begin{equation}
  \frac{1}{2\pi \ls}  \int_{\Sigma^N_{\pm}}F_2=\frac{1}{2\pi \ls}  \int_{\Sigma^S_{\pm}}F_2\equiv M\, ,
\end{equation}
where we have used that all two-cycles are homologous $[\Sigma^N_{\pm}]=[\Sigma^S_{\pm}]$. Consider now a four-cycle constructed by fibring a (trivial) two-cycle over the bolt. We will need to compute the first Chern class of the normal line bundle to each facet. To do this we need to compute the dual basis for each fixed point on the resolved $T^{1,1}$. 
For the cones $\sigma_1^\pm=\{v_1,v_2,v_3\}$ the dual bases are
\begin{align}
    u_1^{+}&=(1,0,-1)\, ,\quad u_2^+=(0,0,1)\, ,\quad u_3^+=(0,1,1)\, ,\\
    u_1^-&=(1,-1,0)\, ,\quad u_3^-=(0,0,1)\, ,\quad u_4^-=(0,1,-1)\, .
\end{align}
Therefore on the cone $\tau^{\pm}$ the normal bundle to the facet defined by the vector $v_a$ is 
\begin{equation}
    c_1(L_a)=\sum_{i=1}^{3}(u_a^{\pm})_{i}c_1(\mathcal{L}_i)\,,
\end{equation}
where the result depends on which fixed point one is looking at.
Using this we can compute the integral of the four-form flux through the trivial four-cycles, and we find the two constraints:
\begin{equation}
    \int_{\Sigma^S_{+}}\Phi_2=  \int_{\Sigma^N_{+}}\Phi_2+ 2 \pi \ls^3M \y_{123}^-\, ,\quad  \int_{\Sigma^S_{-}}\Phi_2=  \int_{\Sigma^N_{-}}\Phi_2+ 2 \pi \ls^3M \y_{134}^-\, ,
\end{equation}
where $\y_{123}^-$ and $\y_{134}^-$ are defined in \eqref{eq:ypmT11}.

Finally, we may consider the six-cycles $[\hat{D}_a]$ consisting of the four-cycles $[D_a]\subset X_6$ fibred over the Riemann surface. In the blown-down solution the four-cycles are constructed by fibring the three-cycle divisors of $T^{1,1}$ over an interval. The six-cycles obtained by fibring these four-cycles over the Riemann surface are not independent and instead satisfy\footnote{This can be obtained by performing a similar analysis as in appendix \ref{app:homology}. Alternatively, one can use the results in appendix B of \cite{Gauntlett:2019roi} for Sasaki--Einstein fibres and introduce the line interval for the suspension. }
\begin{equation}
    \sum_{a=1}^{4}v_a^{i}[\hat{D}_a]=p_i [X_6]\, .
\end{equation}
Imposing the homology relation allows us to fix 
\begin{align}
    \int_{\Sigma^N_-}\Phi_2&=\frac{(b_1-b_2)b_3}{b_2(b_1-b_3)}\int_{\Sigma^N_+}\Phi_2+\frac{(\y_{134}^-)^2\ls^3n_0 \pi^2\vol(\vec{b})}{12}\bigg[\frac{(b_2-b_3)b_3(b_1^2-b_2b_3)}{b_1(b_1-b_3)}p_1\nonumber\\
    &+\frac{(b_1-2b_2)(b_2-b_3)b_3}{b_2}p_2+\frac{(b_1-b_2)(b_1-2b_3)(b_2-b_3)}{b_1-b_3}p_3\bigg]\, ,
\end{align}
where we have presented the $M=0$ case for simplicity only. 
Upon inserting this together with the other constraints into the action, it remains to extremize over $Z_0$ which fixes
\begin{equation}
    Z_0=-\frac{M \varepsilon}{m}\, ,
\end{equation}
and the partially on-shell action takes the universal form
\begin{equation}
    I=\frac{9}{4}F_{S^3}^{T^{1,1}}\Big(\sum_{i=1}^{3}p_i\partial_{b_i}\mathcal{F}-\frac{m}{\varepsilon}\mathcal{F}\Big)+\frac{\ii\pi N M^2}{m n_0}\, ,
\end{equation}
where for $m=0$ we must set $M=0$. The prepotential is
\begin{equation}
    \mathcal{F}=\bigg(\frac{\vol_{T^{1,1}}}{\vol_{T^{1,1}}(\vec{b}\,)}\bigg)^{2/3}\, ,
\end{equation}
where $\vol_{T^{1,1}}(\vec{b}\,)$ is the Sasakian volume.

%%%%%%%%%%%%%%%%%%%%%%%%%%%%%%%%%%%%%%%%%%%
%%%           
%%%%%%%%%%%%%%%%%%%%%%%%%%%%%%%%%%%%%%%%%%%

\subsection{General toric Sasaki--Einstein manifolds}

The localization computation presented above for the suspension of $T^{1,1}$ extends naturally to an arbitrary toric Sasaki--Einstein five-manifold.
Let $Y_5$ be a toric Sasaki--Einstein manifold with Calabi--Yau cone $C(Y_5)$. The cone is specified by a convex lattice polygon, or equivalently by primitive generators
\begin{equation}
v_a=(1,w_a)\in\mathbb{Z}^3\,,\qquad a=1,\ldots,d\, ,
\end{equation}
where $d$ is the number of facets of the toric diagram.

In general the cone is singular at the tip, as in the $T^{1,1}$ example above. To perform our equivariant localization we choose a simplicial crepant subdivision of the fan, or equivalently a triangulation of the toric diagram using its lattice points. The resulting toric variety is a crepant partial
resolution of the cone. We do not require the triangulation to be unimodular, and hence allow orbifold singularities. Different choices of triangulation are related by toric flops.

Each maximal cone
\begin{equation}
\tau_{a_1a_2a_3}=\langle v_{a_1},v_{a_2},v_{a_3}\rangle
\end{equation}
corresponds to an isolated fixed point of the torus action. 
One can now follow a similar procedure as above. Each maximal cone contributes one fixed point and if the subdivision is not smooth the local geometry at each fixed point is $\mathbb{C}^3/\Gamma$ where $|\Gamma|=|\text{det}(v_{a_1},v_{a_2},v_{a_3})|$ and one introduces the corresponding orbifold factor in the BV-AB formula. One sets the Page fluxes through the exceptional divisors introduced by the subdivision to zero and at the very end blows the exceptional divisors back down. Since the subdivision is crepant, the canonical bundle is preserved, and the resulting localization formula depends only on the original Sasaki--Einstein geometry and not on the particular choice of simplicial crepant subdivision.

The end result of this procedure is that the partially on-shell action satisfies \eqref{Ifinal4d} where the prepotential is
\begin{equation}
    \mathcal{F}=\bigg(\frac{\vol_{\text{SE}_5}}{\vol_{\text{SE}_5}(\vec{b})}\bigg)^{2/3}\, .
\end{equation}
Here the numerator is the on-shell volume and the denominator is the Sasakian volume, with the latter depending on the three R-symmetry parameters $b_i$.

\bibliographystyle{JHEP}

\bibliography{IIloc}

\end{document}